\documentclass[aps,prd,twocolumn,showpacs,10pt,superscriptaddress,preprintnumbers]{revtex4-1}
\usepackage{graphicx}
\usepackage{amsfonts}
\usepackage{amssymb}
\usepackage{amsmath}
\usepackage{color}
\usepackage{epstopdf}

\newcommand{\beq}{\begin {equation}}
\newcommand{\eeq}{\end   {equation}}
\newcommand{\bea}{\begin {eqnarray}}
\newcommand{\eea}{\end   {eqnarray}}
\newcommand{\baa}{\begin {array}   }
\newcommand{\eaa}{\end   {array}   }
\newcommand{\nn }{\nonumber        }

\begin{document}
\preprint{
	{\vbox {
			\hbox{\bf LA-UR-21-24315}
}}}

\title{Determining Higgs boson width at electron-positron colliders}

\author{Bin Yan}
\email{binyan@lanl.gov}
\affiliation{Theoretical Division, Group T-2, MS B283, Los Alamos National Laboratory, P.O. Box 1663, Los Alamos, NM 87545, USA}

\begin{abstract}
Probing Higgs width $\Gamma_h$ is critical to test  the Higgs properties. In this work we propose to measure $\Gamma_h$ at the $e^+e^-$ collider with a model-independent analysis under the Standard Model Effective Field Theory framework. We demonstrate that making use of the cross section measurements from $e^+e^-\to Zh$, $e^+e^-\to \nu_e\bar{\nu}_eh$ production and Higgs decay branching ratios  $h\to WW^*/ZZ^*$, one could determine $\Gamma_h$ at a percentage level with a center of mass energy $\sqrt{s}=250$ and 350 GeV and integrated luminosity $5~{\rm ab}^{-1}$. This conclusion would not depend on the assumption of the fermion Yukawa interactions.  We further apply this result to constrain the fermion Yukawa couplings and it shows that the couplings could be well constrained.

\end{abstract}

\maketitle

\section{Introduction}
After the discovery of the Higgs boson at the Large Hadron Collider (LHC),  precision measurements of the properties of the Higgs  at the LHC and future colliders have become a prior task of particle physics.  Determining the couplings of the Higgs boson to particles in the Standard Model (SM) is one of the avenues to verify the SM and search for the possible new physics (NP) beyond the SM (BSM). Applying the narrow width approximation for the Higgs production and decay, the scattering rate of process $i\to h \to f$ could be factorized as the Higgs production cross section and decay branching ratio, i.e.,
\bea
\sigma_{i\to h \to f}=\sigma(i\to h){\rm BR}(h\to f)\propto\dfrac{g_i^2g_f^2}{\Gamma_h},
\eea
where $g_{i}~(g_f)$ is the Higgs coupling from the initial (final) state and $\Gamma_h$ is Higgs width. It is evident that any  attempts to extract  the information of $g_{i,f}$, one needs an assumption of $\Gamma_h$.   Therefore,   measuring the Higgs width with a model independent method becomes crucial for us to understand the Higgs properties.  However,  $\Gamma_h\sim4~{\rm MeV}$ in the SM for 125 GeV Higgs boson,   it would be a challenge to probe the Higgs width directly with a desirable accuracy at the LHC and future colliders due to the limitation of the detector energy and momentum resolution of the final states. Alternatively, $\Gamma_h$ could be probed indirectly at the LHC with additional theoretical assumptions.   For example, $\Gamma_h$ could be obtained by (1) comparing the production rates  of on-shell and off-shell Higgs production~\cite{Caola:2013yja,Campbell:2013una,Campbell:2013wga}; (2) the invariant mass distribution of $\gamma\gamma$ and $ZZ$ from  the interference between the Higgs production and the continuum background~\cite{Dixon:2013haa,Campbell:2017rke}; (3) $t\bar{t}h$ and $t\bar{t}t\bar{t}$ production rates~\cite{Cao:2016wib,Cao:2019ygh}.
So far the  ATLAS~\cite{Aad:2015xua,Aaboud:2018puo} and CMS ~\cite{Khachatryan:2014iha,Khachatryan:2016ctc,CMS:2018bwq} collaborations  give  the upper bound of $\Gamma_h\leq 14.4~{\rm MeV}$ at 95\%  confidence level based on the first method. 

Due to  the clean and readily identifiable signature of the Higgs boson  at the $e^+e^-$ collider, we expect the accuracy of  Higgs width could be much improved at the lepton collider. There are three major proposals for the lepton colliders, the Circular Electron Positron Collider (CEPC)~\cite{CEPCStudyGroup:2018rmc}, the Future  Circular Collider (FCC-ee)~\cite{Gomez-Ceballos:2013zzn}, and the International Linear Collider (ILC)~\cite{Baer:2013cma}. At the $e^+e^-$ collider, $\Gamma_h$ could  be probed indirectly with a high accuracy by the measurements of Higgs  production rates and the decay branching ratios~\cite{Han:2013kya,Asner:2013psa,CDR,Durig:2014lfa,Chen:2016prx,Barklow:2017suo,Lafaye:2017kgf}. For example, $\Gamma_h$ could be measured through the $e^+e^-\to Zh (\to ZZ^*)$ production channel, i.e.,
\bea
\Gamma_h=\dfrac{\Gamma(h\to ZZ^*)}{BR(h\to ZZ^*)}\sim \dfrac{\sigma(e^+e^-\to Zh)}{BR(h\to ZZ^*)},
\label{eq:gamma}
\eea
where $\Gamma(h\to ZZ^*)$ ($BR(h\to ZZ^*)$) is the partial decay width  (branching ratio) of $h\to ZZ^*$, and $\sigma(e^+e^-\to Zh)$ is the production rate of $e^+e^-\to Zh$. 
To determine $\Gamma_h$ via this strategy, it depends on the $\kappa$-framework assumption on the Higgs couplings; i.e.  all the Higgs couplings are SM-like and the deviations are dressed by one scale factors $\kappa_i$ for the coupling of particle $i$ to Higgs boson.  However, an important feature of the $\kappa$ framework is that the kinematics of the Higgs boson are same as the SM. Going beyond $\kappa$-framework  becomes important for the NP which  has the different Lorentz structures of the Higgs couplings compared to the SM, e.g. the BSM operators under the SM effective field theory (SMEFT).
In that case, the presence of the new $hZZ$ anomalous couplings will ruin the strategy in Eq.~\ref{eq:gamma}.  To overcome this  problem, we need a separate method to  determine the size of each operators. 
In this work, we try to present the minimum number of observabels that are needed to extract $\Gamma_h$ at the $e^+e^-$ collider under the SMEFT framework.  We argue that $\Gamma_h$ could be determined via combining the data from the cross sections  of $e^+e^-\to Zh$, $e^+e^-\to \nu_e\bar{\nu}_e h$ productions and branching ratios  of $h\to WW^*/ZZ^*$. We emphasize that our strategy would rely on the Higgs gauge couplings alone, while not for the assumption of the Yukawa interactions. It shows that the accuracy of $\Gamma_h$ could be reached at percentage level at the CEPC and the result is comparable to the method in Eq.~\ref{eq:gamma}~\cite{Chen:2016prx}.

\section{Higgs electroweak gauge couplings}
\label{sec:EFT}
Given the null results so far for NP searches at the LHC, the SMEFT is perfectly applicable at the future lepton colliders with center-of-mass energy $\sqrt{s}<1~{\rm TeV}$. 
The NP effects under the SMEFT could be parameterized by a set of higher dimensional operators  which are invariant under the Lorentz group and gauge symmetry $SU(3)_c\otimes SU(2)_L\otimes U(1)_Y$~\cite{Buchmuller:1985jz,Giudice:2007fh,Grzadkowski:2010es,Li:2020gnx},
\bea
\mathcal{L}_{\rm eff}=\mathcal{L}_{\rm SM}+\sum_i(c_i\mathcal{O}_i+h.c.)+...,
\eea
where $\mathcal{L}_{\rm SM}$ denotes the SM Lagrangian; $c_i$ is the Wilson coefficient of the dimension-6 operator $\mathcal{O}_i$ and the dots denote higher dimension operators which will be ignored in this work. The operators that contribute to the Higgs gauge couplings in the SILH basis are~\cite{Giudice:2007fh},
\begin{align}
\mathcal{O}_H&=\dfrac{1}{2v^2}\partial^\mu(H^+H)\partial_\mu(H^+H), \nn\\
\mathcal{O}_{BB} &=\dfrac{g^{\prime 2}}{m_W^2}H^+HB_{\mu\nu}B^{\mu\nu},\nn\\
\mathcal{O}_{HB} &=\dfrac{i g^\prime}{m_W^2}(D^\mu H)^+(D^\nu H)B_{\mu\nu}, \nn\\
\mathcal{O}_{HW} &=\dfrac{ig}{m_W^2}(D^\mu H)^+\sigma^i(D^\nu H)W_{\mu\nu}^i,\nn\\
\mathcal{O}_W &= \dfrac{ig}{2 m_W^2} D^\nu W_{\mu\nu}^i (H^+  \sigma^i  {\overleftrightarrow { D^\mu}} H) ,\nn\\
\mathcal{O}_B &= \dfrac{i g^\prime}{2 m_W^2} \partial^\nu B_{\mu\nu} (H^+  {\overleftrightarrow { D^\mu}} H),\nn\\
\mathcal{O}_T &=\dfrac{1}{2v^2} (H^+  {\overleftrightarrow { D^\mu}} H) (H^+  {\overleftrightarrow { D_\mu}} H),
\label{eq:dim6}
\end{align}
where $D_\mu=\partial_\mu-ig (\tau^i/2)W_\mu^i-ig^\prime Y B_\mu$ is the gauge covariant derivative, $g$ and $g^\prime$ are the gauge couplings of $SU(2)_L$ and $U(1)_Y$; $Y$ is the hypercharge of the field.  Note  $H^+  {\overleftrightarrow { D^\mu}} H\equiv H^+D_\mu H-(D_\mu H)^+H$ and $H$ is the $SU(2)_L$ weak doublet of the  Higgs field, $W_{\mu\nu}^i$ and $B_{\mu\nu}$ are the gauge boson field strength tensor of $SU(2)_L$ and $U(1)_Y$, respectively.  
The operators $\mathcal{O}_{W,B,T}$ are constrained strongly by the current electroweak precision measurements~\cite{Giudice:2007fh} and the bounds would be strengthened in the future lepton colliders~\cite{DeBlas:2019qco}, as a result, they will be neglected in this work. 
After the electroweak symmetry breaking $\langle H \rangle=v/\sqrt{2}$ with $v=246 ~{\rm GeV}$,  above operators generate the following  effective couplings  of Higgs to the gauge bosons,
\begin{align}
\mathcal{L}_{hVV}&=\dfrac{h}{v}\left[g_{h\gamma\gamma}A_{\mu\nu}A^{\mu\nu}+g_{hZ\gamma}^{a}Z^\mu\partial^\nu A_{\mu\nu}+g_{hZ\gamma}^{b}A_{\mu\nu}Z^{\mu\nu}\right.\nn\\
&+g_{hZZ}^{a}Z^\mu\partial^\nu Z_{\mu\nu}+g_{hZZ}^{b}Z_{\mu\nu}Z^{\mu\nu}+g_{hZZ}^{c}Z_\mu Z^\mu\nn\\
&+g_{hWW}^{a}(W^{-\mu}\partial^\nu W_{\mu\nu}^+ +h.c.)+g_{hWW}^{b}W_{\mu\nu}^+W^{-\mu\nu}\nn\\
&\left.+g_{hWW}^{c}W_\mu^+W^{-\nu}\right],
\label{eq:hvv}
\end{align}
where $V_{\mu\nu}=\partial_\mu V_\nu-\partial_\nu V_\mu$, with $V=A,~Z,~W^{\pm}$.  The coefficients of above effective couplings are related to the Wilson coefficients of the dimension-6 operators as follows,
\begin{align}
&g_{h\gamma\gamma}=4c_{BB} s_W^2;&
&g_{hZ\gamma}^a=2t_W(c_{HW}-c_{HB});\nn\\
&g_{hZ\gamma}^b=t_W(c_{HB}-c_{HW}-8s_W^2c_{BB});&
&g_{hZZ}^a =2t_W^2 c_{HB}+2c_{HW};\nn\\
&g_{hZZ}^b =4t_W^2s_W^2c_{BB}-t_W^2 c_{HB}-c_{HW};&
&g_{hZZ}^c =(1-\dfrac{c_H}{2})m_Z^2;\nn\\
&g_{hWW}^a =2 c_{HW};&
&g_{hWW}^b =-2 c_{HW};\nn\\
&g_{hWW}^c =2(1-\dfrac{c_H}{2})m_W^2.
\label{eq:chvv}
\end{align}
Here  $s_W\equiv \sin\theta_W$ and $t_W=\tan\theta_W$ with $\theta_W$ is the weak mixing angle.
In the following, we will discuss the Higgs production and decay branching ratios under the general effective Lagrangian (see Eq.~\ref{eq:hvv}) at the $e^+e^-$ collider. A  systematic study on the sensitivities of probing the Higgs couplings at the $e^+e^-$ collider under the SMEFT framework could be found in Refs.~\cite{Ge:2016zro,Chiu:2017yrx,Khanpour:2017cfq,Durieux:2017rsg,Barklow:2017suo,Cao:2018cms,Xie:2021xtl}. We should note that the operators which are related to the SM fermions may also contribute to the observables of we are considering, however, it is beyond the scope of this paper and could be found in Refs.~\cite{Ge:2016zro,Chiu:2017yrx,Khanpour:2017cfq,Durieux:2017rsg}.

 \section{The observables}
 \label{sec:obs}
 \subsection{Higgs boson production cross sections}
 \begin{figure}
 \begin{center}
\includegraphics[scale=0.3]{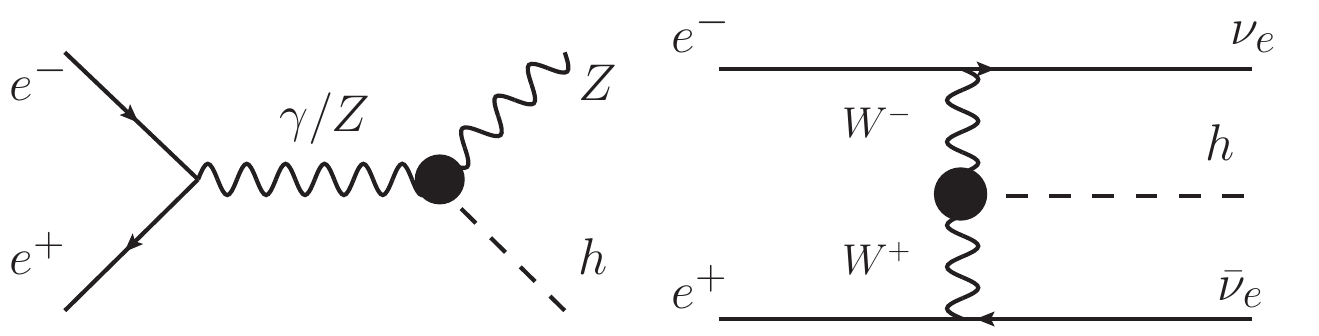}
\caption{Illustrative Feynman diagrams of $e^+e^-\to h Z$ and $e^+e^-\to \nu_e\bar{\nu}_e h$. The black dots denote the effective couplings including the new physics effects. }
\label{fig:feyman}
\end{center}
\end{figure} 
Next we discuss the cross sections of processes $e^+e^-\to Zh$ ($\sigma_{Zh}$) and $e^+e^-\to \nu_e\bar{\nu}_e h$ ($\sigma_{\nu\bar{\nu}h}$) at the lepton collider.  
The generic $hZZ$, $hZ\gamma$  and $hW^+W^-$ anomalous couplings in Eq.~\ref{eq:hvv} could contribute to the cross sections $\sigma_{Zh}$ and $\sigma_{\nu\bar{\nu}h}$;  see Fig.~\ref{fig:feyman}. The excellent agreement between the SM and data indicates that deviations from the NP should be small. Hence, we restrict ourselves to the interference terms between SM and the BSM operators, i.e. the leading order of the coefficients $c_i$.
The total cross sections can be written as a linear combination of the SM contribution and NP corrections,
\begin{align}
&\sigma_{Zh}=\sigma_{Zh}^{\rm SM}\left(1-c_H+\sum_{i,\mathbb{V}} g_{hZ\mathbb{V}}^iR_{Z\mathbb{V}}^i\right),\nn\\
&\sigma_{\nu\bar{\nu}h}=\sigma_{\nu\bar{\nu}h}^{\rm SM}\left(1-c_H+\sum_{i} g_{hWW}^iR_{WW}^i\right),
\label{eq:sigma}
\end{align}
where $g_{hZ\mathbb{V}}^i$ and $g_{hWW}^i$,  with $i=a, b$ and $\mathbb{V}=\gamma,Z$ are the effective couplings between Higgs and gauge bosons; see Eq.~\ref{eq:chvv}. $\sigma_{Zh}^{\rm SM}$ and $\sigma_{\nu\bar{\nu}h}^{\rm SM}$  are the production cross sections of $e^+e^-\to Zh$ and $e^+e^-\to \nu_e\bar{\nu}_eh$ in the SM , respectively.  The coefficients $R_{Z\mathbb{V}}^i$ and $R_{WW}^i$ describe the interference effects between the SM and the Higgs anomalous couplings and their  values depend on the collider energy ($\sqrt{s}$).  The coefficients $R_{Z\mathbb{V}}^i$ can be calculated  with analytical method and the results are,
\begin{align}
R_{ZZ}^a&=1+\frac{s}{m_Z^2},&
R_{ZZ}^b &=\frac{12E_Z\sqrt{s}}{2m_Z^2+E_Z^2},\nn\\
R_{Z\gamma}^a&=(1-\frac{s}{m_Z^2})F_{Z\gamma},&
R_{Z\gamma}^b&=\frac{6E_Z(m_Z^2-s)}{\sqrt{s}(2m_Z^2+E_Z^2)}F_{Z\gamma},
\end{align}
where $E_Z$ is  energy of the $Z$ boson in the center-of-mass frame and $F_{Z\gamma}$ is the coupling ratio,
\begin{align}
E_Z&=\frac{s+m_Z^2-m_h^2}{2\sqrt{s}},&
F_{Z\gamma}&=\dfrac{e(g_L^e+g_R^e)}{(g_L^e)^2+(g_R^e)^2}\simeq -0.18.
\end{align}
Here $m_{Z}$ and $m_h$ are the $Z$ boson and Higgs boson mass, respectively. $e$ is the electron charge; $g_L^e=g/c_W(-1/2+s_W^2)$ and $g_R^e=g/c_W s_W^2$  with $c_W=\cos\theta_W$ are the left- and right-handed gauge couplings of the $Z$ boson to the electron.  
The  analytical results for the coefficients $R_{WW}^i$ are not available, thus we will show the numerical results only in this work. 

\begin{figure}
\begin{center}
\includegraphics[scale=0.3]{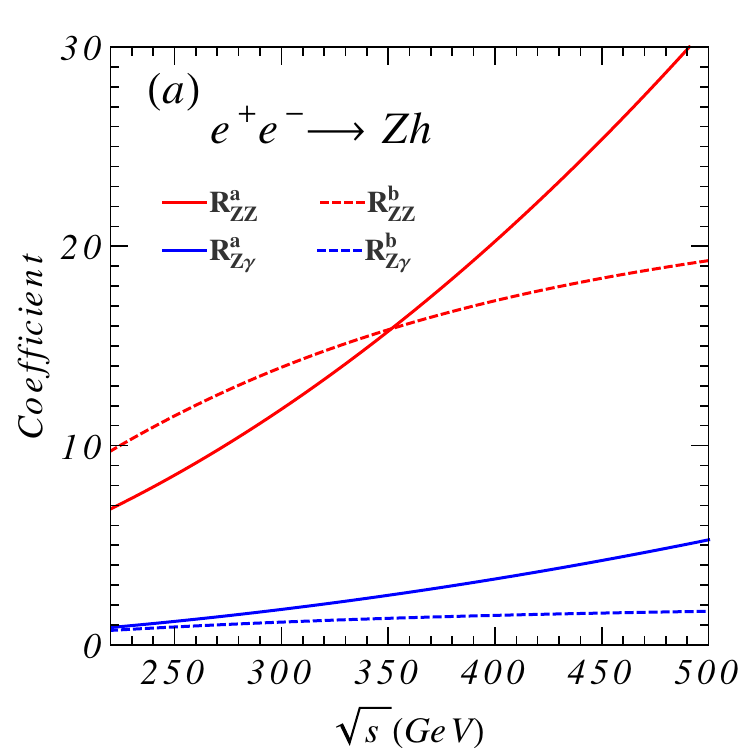}
\includegraphics[scale=0.3]{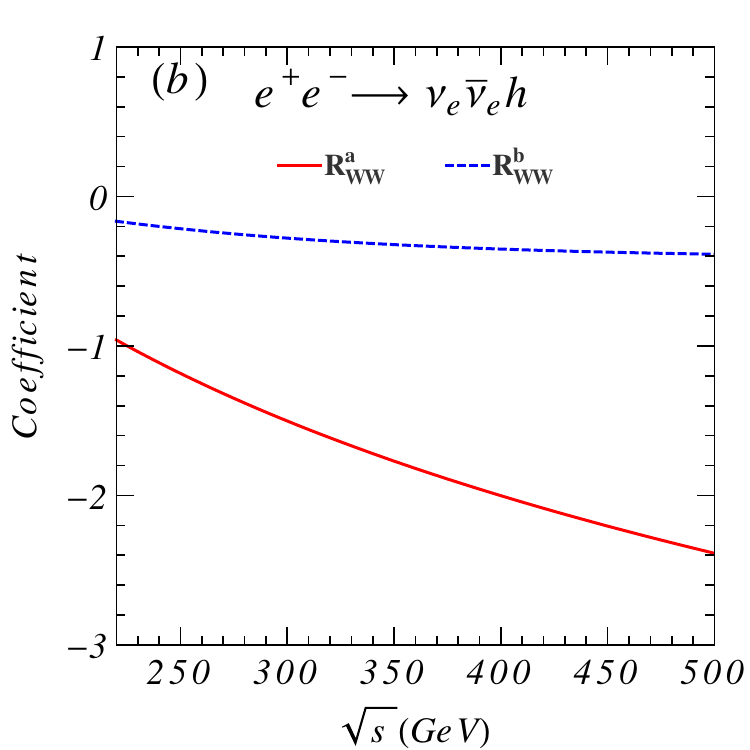}
\caption{Dependence on the collider energy $\sqrt{s}$ of the coefficients: (a)  $e^+e^-\to h Z$ and (b) $e^+e^-\to \nu_e\bar{\nu}_e h$. }
\label{fig:Rcs}
\end{center}
\end{figure}

Figure~\ref{fig:Rcs} displays the coefficients $g_{hZ\mathbb{V}}^i$ and $g_{hWW}^i$  as a function of the collider energy $\sqrt{s}$.
Obviously, $R_{ZZ}^a$  (red solid line) is much sensitive to the collider energy than $R_{ZZ}^b$ (red dashed line), and  $R_{Z\gamma}^a$ (blue solid line) has a similar energy dependence as $R_{ZZ}^a$, but its value is highly suppressed by the coupling ratio $F_{Z\gamma}$; see Fig.~\ref{fig:Rcs}(a). For the $e^+e^-\to \nu_e\bar{\nu}_e h$ production,  the absolute value of the coefficient $R_{WW}^a$ (red solid line) is much larger  than  $R_{WW}^b$ (blue dashed line) and it also  shows a stronger energy dependence compared to $R_{WW}^b$. It arises from the fact that the matrix element of $R_{WW}^a$ is proportional to the momentum transfer $t_1=(k_{\nu_e}-k_{e^-})^2$ and $t_2=(k_{\bar{\nu}_e}-k_{e^+})^2$, where $k_i$ is the momentum of the particle $i$.  As a result, the $R_{WW}^a$ could be enhanced  when the momentum $k_{\nu_e/\bar{\nu_e}}$ is antiparallel to the $k_{e^-/e^+}$.

The cross section $\sigma_{Zh}$ can be measured at the $e^+e^-$ collider with the recoil mass method by tagging the decay products of the associated $Z$ boson and the result is  independently of the Higgs decay. However, the direct measurement of $\sigma_{\nu\bar{\nu}h}$ is relying on the assumption of the Higgs decay branching ratios.  Alternatively, we can extract  $\sigma_{\nu\bar{\nu}h}$ from the ratio of the cross sections of $e^+e^-\to Zh$ and $e^+e^-\to \nu_e\bar{\nu}_e h$  processes with one specific Higgs decay mode.
The advantage of this observable is that the $\sigma_{\nu\bar{\nu}h}$ could be measured without the assumption of the Higgs decay. As an example, we will focus on the $b\bar{b}$ mode since both the $Zh$ and $ \nu_e\bar{\nu}_eh$ production with $h\to b\bar{b}$ could be measured very well at the future lepton colliders~\cite{CDR,CEPCStudyGroup:2018ghi}.  The ratio is defined as, 
\begin{align}
R_{\rm cs}&=\dfrac{\sigma_{Zh,h\to b\bar{b}}/\sigma_{Zh,h\to b\bar{b}}^{\rm SM}}{\sigma_{\nu\bar{\nu}h,h\to b\bar{b}}/\sigma_{\nu\bar{\nu}h,h\to b\bar{b}}^{\rm SM}}=\frac{R_{Zh}}{R_{\nu\bar{\nu}h}},
\label{eq:ratio}
\end{align}
where $R_{Zh}=\sigma_{Zh}/\sigma_{Zh}^{\rm SM}$, $R_{\nu\bar{\nu}h}=\sigma_{\nu\bar{\nu}h}/\sigma_{\nu\bar{\nu}h}^{\rm SM}$ and the uncertainty from the unknown $hb\bar{b}$ coupling and Higgs width are cancelled.

\subsection{Higgs decay branching ratios}
 \begin{figure}
 \begin{center}
\includegraphics[scale=0.27]{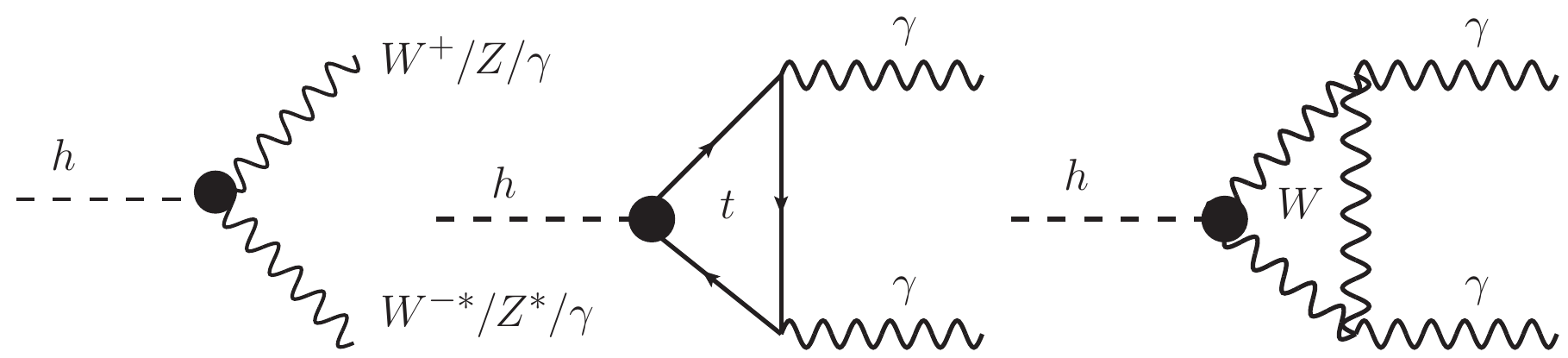}
\caption{Illustrative Feynman diagrams of $h \to W^+W^{-*}/ZZ^*/\gamma\gamma$. The black dots denote the effective couplings including the new physics effects. }
\label{fig:hdecay}
\end{center}
\end{figure} 

The operators in Eq.~\ref{eq:dim6} will also change the partial decay widths of Higgs to gauge bosons~\footnote{The Higgs width under the general BSM operators could be found in Ref.~\cite{Brivio:2019myy}}. In this work, we will focus on $h\to ZZ^*/WW^{*}/\gamma\gamma$ modes; see Fig.~\ref{fig:hdecay}. The partial decay widths of Higgs to  $ZZ^*$ and $WW^*$ can be expanded as follows,
\begin{align}
\Gamma_{ZZ^*}&=\Gamma_{ZZ^*}^{\rm SM}(1-c_H)+\sum_{i,\mathbb{V}}g_{hZ\mathbb{V}}^i\Gamma_{Z\mathbb{V}}^i,\nn\\
\Gamma_{WW^*}&=\Gamma_{WW^*}^{\rm SM}(1-c_H)+\sum_{i}g_{hWW}^i\Gamma_{WW}^i.
\end{align}
The $hZ\gamma$  anomalous couplings could contribute to $ZZ^*$ mode by $h\to Z+\gamma^*(\to  f\bar{f})$.
The $\Gamma_{ZZ^*}^{\rm SM}$ and $\Gamma_{WW^*}^{\rm SM}$ are the partial decay widths of  $h\to ZZ^*$ and $h\to WW^*$ in the SM, respectively, and
\bea
\Gamma_{VV^*}^{\rm SM}=\frac{g^2g_{\rm effV}^fm_h}{1536\pi^3}\dfrac{m_V^2}{m_W^2}F(\epsilon_V),
\label{eq:gammavv}
\eea
where 
\begin{align}
g_{\rm effV}^f&=\sum_f\left[ (g_{LV}^f)^2+(g_{RV}^f)^2\right]N_c^f.
\end{align}
Here  $g_{LV/RV}^f$ are the left- and right-handed gauge couplings of the fermion $f$ to the gauge boson $V=W,Z$; $N_c^f=1$ for the leptons and $N_c^f=3$ for the quarks; $\epsilon_V=m_V/m_h$. After combining all possible final states, we obtain  the effective couplings
\begin{align}
g_{\rm eff Z}^f&=\dfrac{g^2}{12c_W^2}(160s_W^4-120s_W^2+63)\simeq 2.03,\nn\\
g_{\rm eff W}^f&=9g^2\simeq 3.85.
\end{align}
The function $F(\epsilon)$ in Eq.~\ref{eq:gammavv} is~\cite{Rizzo:1980gz,Keung:1984hn},
\begin{align}
F(\epsilon)&=\frac{3(1-8\epsilon^2+20\epsilon^4)}{\sqrt{4\epsilon^2-1}}\arccos\left(\frac{3\epsilon^2-1}{2\epsilon^3}\right)\nn\\
&-(1-\epsilon^2)\left(\frac{47}{2}\epsilon^2-\frac{13}{2}+\frac{1}{\epsilon^2}\right)
-3(1-6\epsilon^2+4\epsilon^4)\ln\epsilon.
\end{align}
The partial decay widths from BSM operators are
\begin{align}
&\Gamma_{ZZ}^a=\dfrac{g^2g_{\rm effZ}^fm_h}{1536\pi^3}\dfrac{1}{\epsilon_W^2}\left[\epsilon_Z^2F(\epsilon_Z)+F_{VV}^a(\epsilon_Z)\right];\nn\\
&\Gamma_{ZZ}^b=\dfrac{g^2g_{\rm effZ}^fm_h}{1536\pi^3}\dfrac{24\epsilon_Z^2}{\epsilon_W^2} F_{VV}^b(\epsilon_Z);\nn\\
&\Gamma_{Z\gamma}^a=\dfrac{g^2g_{\rm eff}^{f\gamma}m_h}{1536\pi^3}\frac{1}{\epsilon_W^2}F_{Z\gamma}^a(\epsilon_Z);\nn\\
&\Gamma_{Z\gamma}^b=\dfrac{g^2g_{\rm eff}^{f\gamma}m_h}{1536\pi^3}\dfrac{12\epsilon_Z^2}{\epsilon_W^2}F_{Z\gamma}^b(\epsilon_Z);\nn\\
&\Gamma_{WW}^a=\dfrac{g^2g_{\rm effW}^fm_h}{1536\pi^3}\dfrac{1}{\epsilon_W^2}\left[\epsilon_W^2F(\epsilon_W)+F_{VV}^a(\epsilon_W)\right];\nn\\
&\Gamma_{WW}^b=\dfrac{g^2g_{\rm effW}^fm_h}{1536\pi^3}12 F_{VV}^b(\epsilon_W).
\label{eq:fun}
\end{align}
The effective coupling $g_{\rm eff}^{f\gamma}$ is defined as 
\bea
g_{\rm eff}^{f\gamma}=\sum_f\left[g_{LZ}^f+g_{RZ}^f\right]Q_f N_c^f e=-\dfrac{5eg}{3c_W}(8s_W^2-3)\simeq 0.46,\nn\\
\eea
where $Q_f$ is the electric charge of the fermion $f$ in unites of $e$. The integration functions in Eq.~\ref{eq:fun} are,
\begin{align}
F_{VV}^a(\epsilon)&=\frac{108\epsilon^6-52\epsilon^4+11\epsilon^2-1}{\sqrt{4\epsilon^2-1}}\arccos\left(\frac{3\epsilon^2-1}{2\epsilon^3}\right)\nn\\
&+\dfrac{1}{6}(\epsilon-1)(\epsilon+1)(179\epsilon^4-100\epsilon^2+17)\nn\\
&+(1-9\epsilon^2+54\epsilon^4-12\epsilon^6)\ln\epsilon;\nn\\
F_{VV}^b(\epsilon)&=\frac{14\epsilon^4-8\epsilon^2+1}{\sqrt{4\epsilon^2-1}}\arccos\left(\frac{3\epsilon^2-1}{2\epsilon^3}\right)
+\frac{9}{2}\epsilon^4-7\epsilon^2\nn\\
&+(-2\epsilon^4+6\epsilon^2-1)\ln\epsilon+\frac{5}{2};\nn\\
F_{Z\gamma}^a(\epsilon)&=(12\epsilon^4-4\epsilon^2+1)\sqrt{4\epsilon^2-1}\arccos\left(\frac{3\epsilon^2-1}{2\epsilon^3}\right)\nn\\
&+\frac{1}{6}(38\epsilon^6-99\epsilon^4+72\epsilon^2-11)
+(36\epsilon^4-6\epsilon^2+1)\ln\epsilon;\nn\\
F_{Z\gamma}^b(\epsilon)&=(2\epsilon^2-1)\sqrt{4\epsilon^2-1}\arccos\left(\frac{3\epsilon^2-1}{2\epsilon^3}\right)\nn\\
&+\frac{3}{2}(\epsilon^2-1)^2+(4\epsilon^2-1)\ln\epsilon.
\end{align}
\begin{figure}
\begin{center}
\includegraphics[scale=0.35]{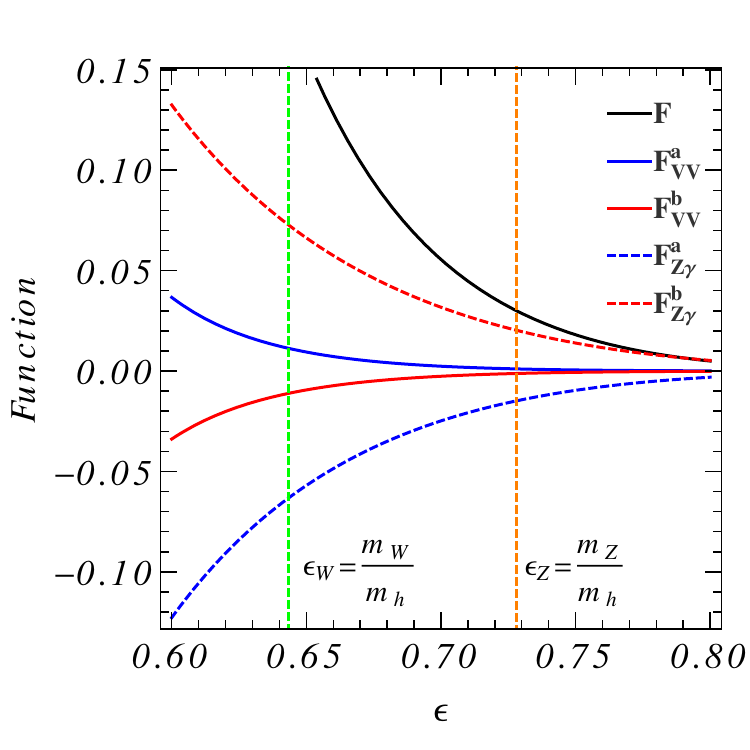}
\caption{The  $\epsilon$ dependence of the integration functions. }
\label{fig:Fun}
\end{center}
\end{figure} 

Figure~\ref{fig:Fun} shows the  $\epsilon$ dependence of functions $F_{VV}^{a,b}$ , $F_{Z\gamma}^{a,b}$ and $F$.  We note that the sign between $F_{VV}^{a(b)}$ (blue solid for $a$ and red solid for $b$) and $F_{Z\gamma}^{a(b)}$ (blue dashed for $a$, red dashed for $b$) is opposite due to the off-shell $W/Z$ propagator. The distribution of $F_{VV}^a$ ($F_{Z\gamma}^a$) also shows a different sign compared to $F_{VV}^b$ ($F_{Z\gamma}^b$).  Such a  behavior could be understood from the couplings in Eq.~\ref{eq:hvv}; i.e.  there is a relative sign in the Feynman rules between the $g_{hVV}^a$ and $g_{hVV}^b$ terms.  Compared to $F_{VV}^{a,b}$,  there is an enhancement effect in $F_{Z\gamma}^{a,b}$ due to the photon propagator. As a result, the  absolute value of $F_{Z\gamma}^{a,b}$ is much larger than $F_{VV}^{a,b}$.  In the limit of $\epsilon\to 1$, Higgs boson can not decay to gauge boson pair with one gauge boson on-shell, so that  all the functions tend to $0$.
For the $Z$ boson, the functions are,
\begin{align}
F(\epsilon_Z)&\simeq0.029,  &F_{VV}^a(\epsilon_Z)&\simeq 0.001,&   F_{VV}^b(\epsilon_Z)\simeq -0.0012,\nn\\
F_{Z\gamma}^a(\epsilon_Z)&\simeq-0.014, &   F_{Z\gamma}^b(\epsilon_Z)&\simeq 0.020.
\end{align}
For the $W$ boson,
\begin{align}
F(\epsilon_W)&\simeq 0.18,& F_{VV}^a(\epsilon_W)&\simeq 0.01,& F_{VV}^b(\epsilon_W)&\simeq-0.01.
\end{align}

The decay mode of Higgs to $\gamma\gamma$ is generated at loop-level in the SM. The contribution from  dimension-6 operators could either come from the tree-level or  at loop level by modifying the couplings in the SM loops. Since the contribution from $hWW$ anomalous couplings in the loop will be highly suppressed, we only consider the SM-like $hWW$ coupling in this decay mode.
The partial decay width of $h\to\gamma\gamma$ is,
\bea
\Gamma_{\gamma\gamma}=\Gamma_{\gamma\gamma}^{\rm SM}\left(1-c_H+\frac{2g_{h\gamma\gamma}}{F_{\gamma\gamma}^{\rm SM}}\right),
\label{eq:Bra}
\eea
where $F_{\gamma\gamma}^{\rm SM}\simeq -0.0046$, induced by the $W$-boson and top quark loops in the SM~\cite{Djouadi:2005gi,Cao:2015iua}.
Note that the $W$-boson loop dominates over the top quark loop, as a result, the possible impact from operator $\mathcal{O}_y=-y_t/v^2H^+H\bar{Q}_L\tilde{H}t_R$ could be ignored.

The branching ratios of $h\to ZZ^*/WW^*/\gamma\gamma$ can be measured by the cross section ratios,
\begin{align}
{\rm BR}_{Z/W/\gamma}=\frac{\sigma_{Zh,h\to ZZ^*/WW^*/\gamma\gamma}}{\sigma_{Zh}}.
\label{eq:Br}
\end{align}
For a given Higgs mass $m_h=125~{\rm GeV}$, the branching ratios could be expressed as follows,
\begin{align}
{\rm BR}_Z &\simeq \frac{\Gamma_{ZZ^*}^{\rm SM}}{\Gamma_h}\left(1-c_H-2.033c_{BB}+2.121c_{HB}+1.868c_{HW}\right);\nn\\
{\rm BR}_W&\simeq \frac{\Gamma_{WW^*}^{\rm SM}}{\Gamma_h}\left(1-c_H+3.787c_{HW}\right);\nn\\
{\rm BR}_\gamma &\simeq \frac{\Gamma_{\gamma\gamma}^{\rm SM}}{\Gamma_h}\left(1-c_H-387c_{BB}\right).
\label{eq:BR}
\end{align}

\subsection{Numerical results}
Next we combine  the measurements of $\sigma_{Zh}$, $\sigma_{\nu\bar{\nu}h}$ and $\rm {BR}_{Z/W/\gamma}$ to determine the Higgs width. Furthermore, we have compared the result of our numerical calculations with that using the MadGraph5~\cite{Alwall:2014hca} and found excellent agreement. 

\begin{figure}
\begin{center}
\includegraphics[scale=0.24]{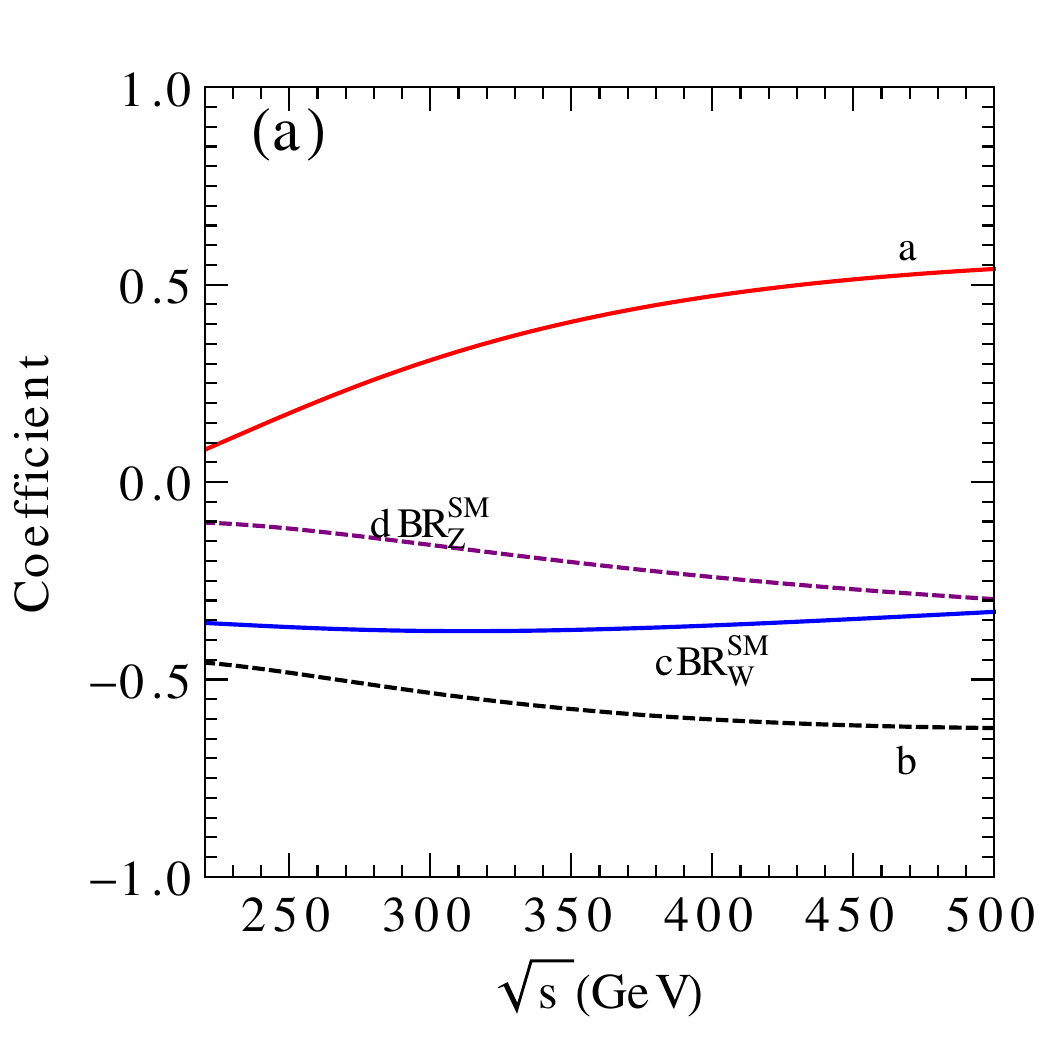}
\includegraphics[scale=0.23]{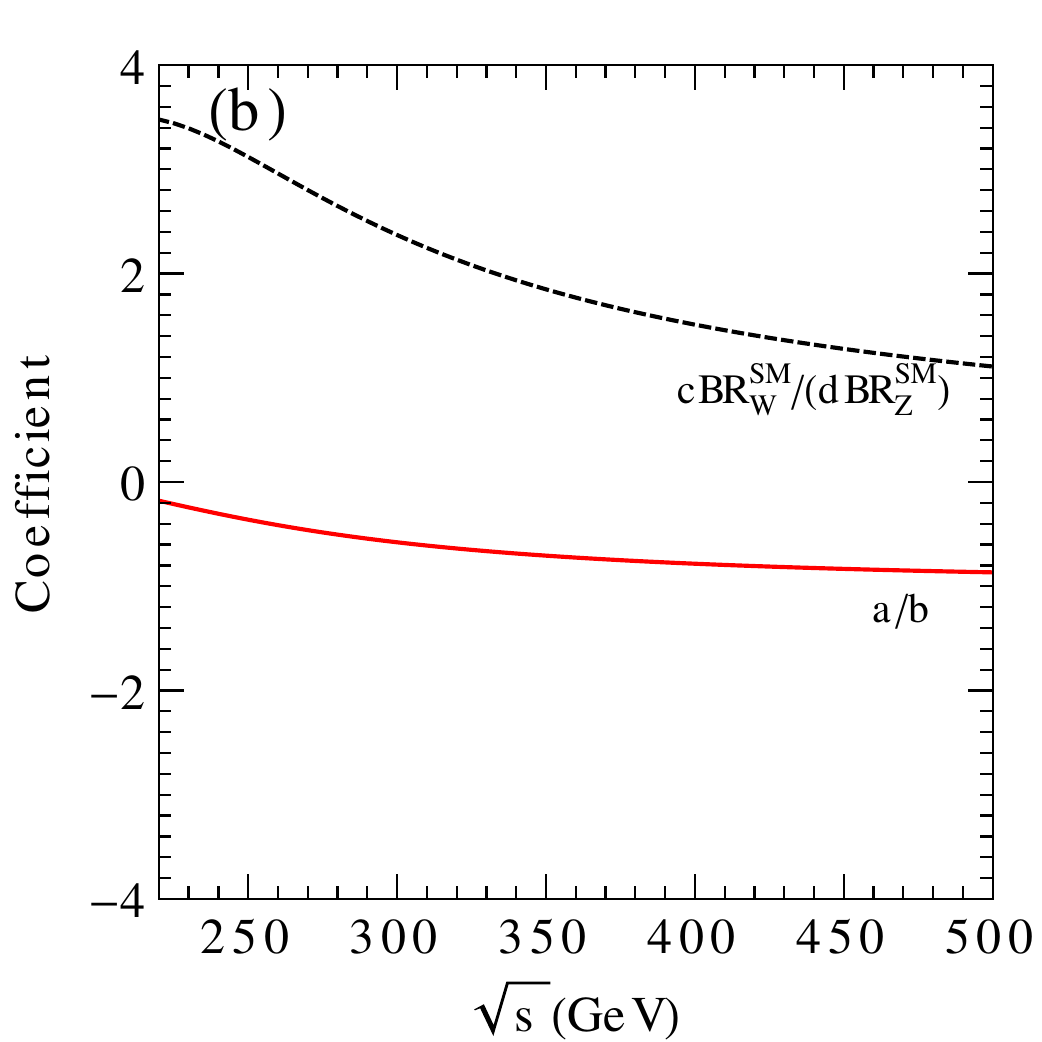}
\caption{Dependence on the collider energy $\sqrt{s}$: (a) the coefficients and (b) the ratios in Eq.~\ref{eq:width}. }
\label{fig:coeff}
\end{center}
\end{figure}

There are five variables in Eqs.~\eqref{eq:sigma},\eqref{eq:ratio},\eqref{eq:BR}, i.e. $c_H,c_{HW},c_{HB},c_{BB},\Gamma_h/\Gamma_h^0$. All of them can be determined by solving the linear equations and it shows that the Higgs width is
\bea
\Gamma_h=\frac{a(R_{cs}-1)+b R_{Zh}}{\rm{BR}_\gamma^{\rm SM} R_\gamma+c {\rm BR}_W^{\rm SM} R_W+d{\rm BR}_Z^{\rm SM} R_Z}\Gamma_h^0,
\label{eq:width}
\eea
where the coefficients $a,b,c,d$ are  dimensionless parameters and their values depend on the collider energy; $\Gamma_h^0=4.07~{\rm MeV}$ is the Higgs width in the SM~\cite{Tanabashi:2018oca} and $R_{\gamma,W,Z}={\rm BR}_{\gamma,W,Z}/{\rm BR}_{\gamma,W,Z}^{\rm SM}$. Due to  $c\rm {BR}_W^{\rm SM}, d\rm{BR}_Z^{\rm SM}\gg \rm{BR}_\gamma^{\rm SM}=2.27\times 10^{-3}$ (see Fig~\ref{fig:coeff}(a)), the Higgs width $\Gamma_h$ depends mainly on $R_{cs/Zh}$ and $\rm{BR}_{W/Z}$. 
We show the energy dependence of the coefficients $a,b,c{\rm BR}_W^{\rm SM},d{\rm BR}_Z^{\rm SM}$ in Fig.~\ref{fig:coeff}(a) and various ratios of those coefficients in Fig.~\ref{fig:coeff}(b). We note that the size of $\Gamma_h$ is sensitive to the cross section $R_{Zh}$ and branching ratio ${\rm BR}_W$ measurements, while $R_{cs}$ and ${\rm BR}_Z$ would become important when $\sqrt{s}>350\sim 400~{\rm GeV}$. The energy dependence of the coefficient $a$ is arise from the fact that the cross section of $e^+e^-\to\nu_e\bar{\nu}_eh$ will be enhanced as the energy increase.

\begin{figure}
\begin{center}
\includegraphics[scale=0.24]{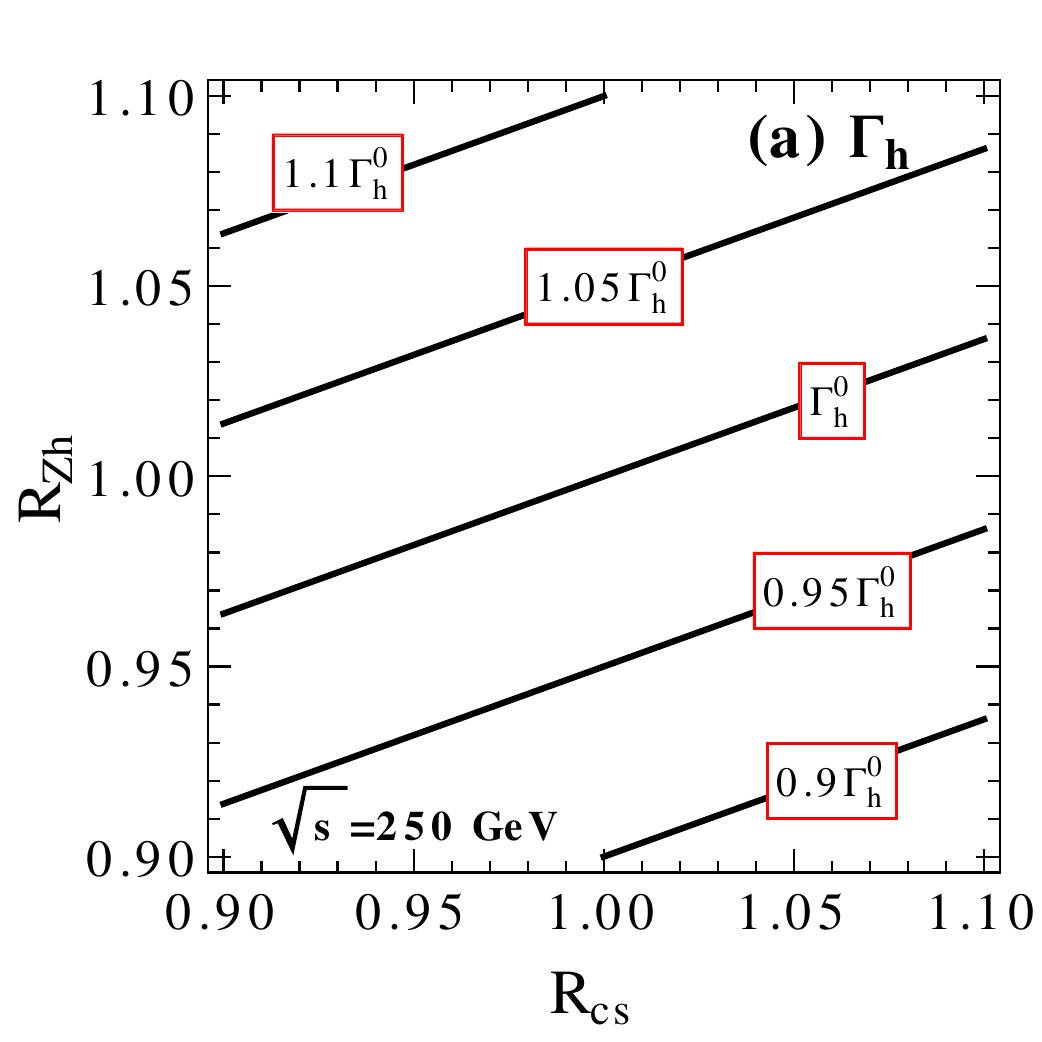}
\includegraphics[scale=0.24]{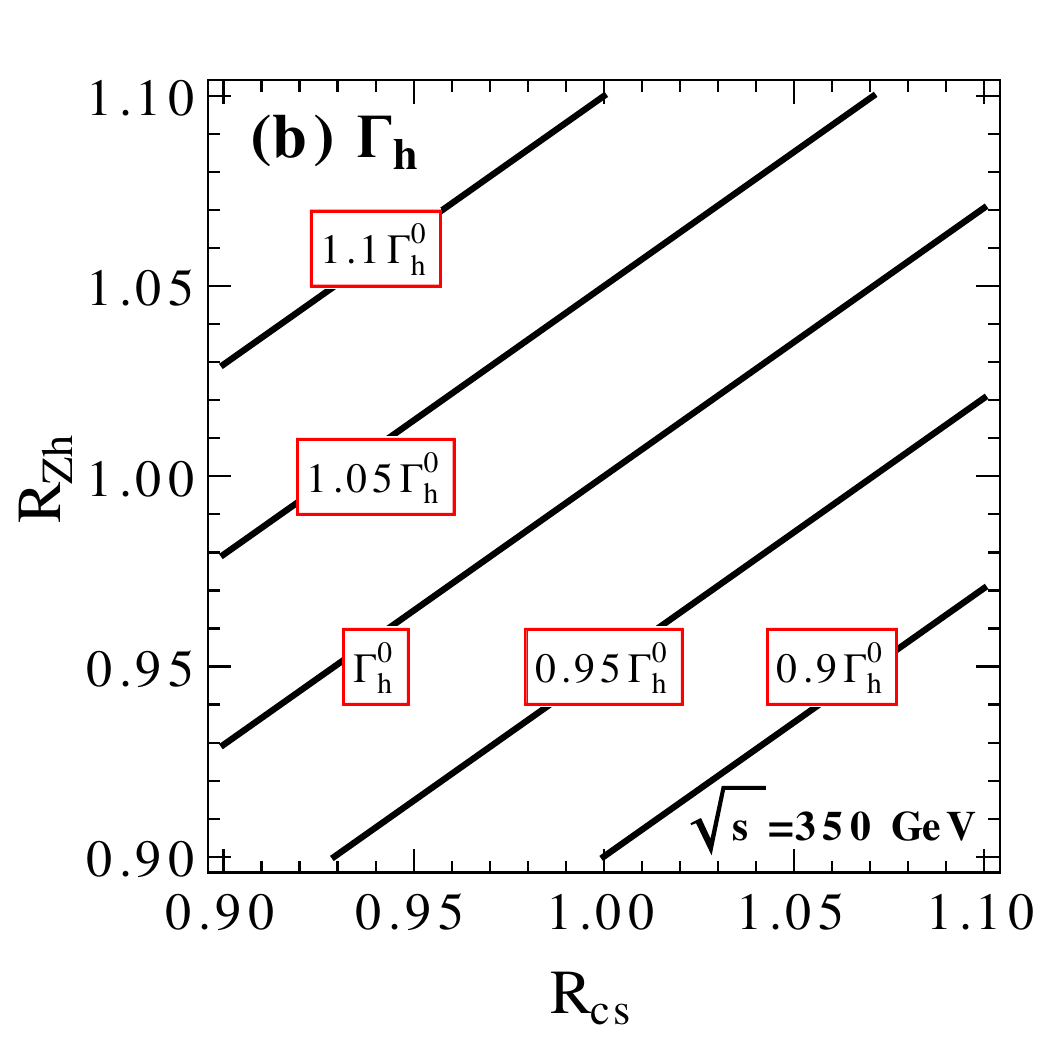}
\includegraphics[scale=0.24]{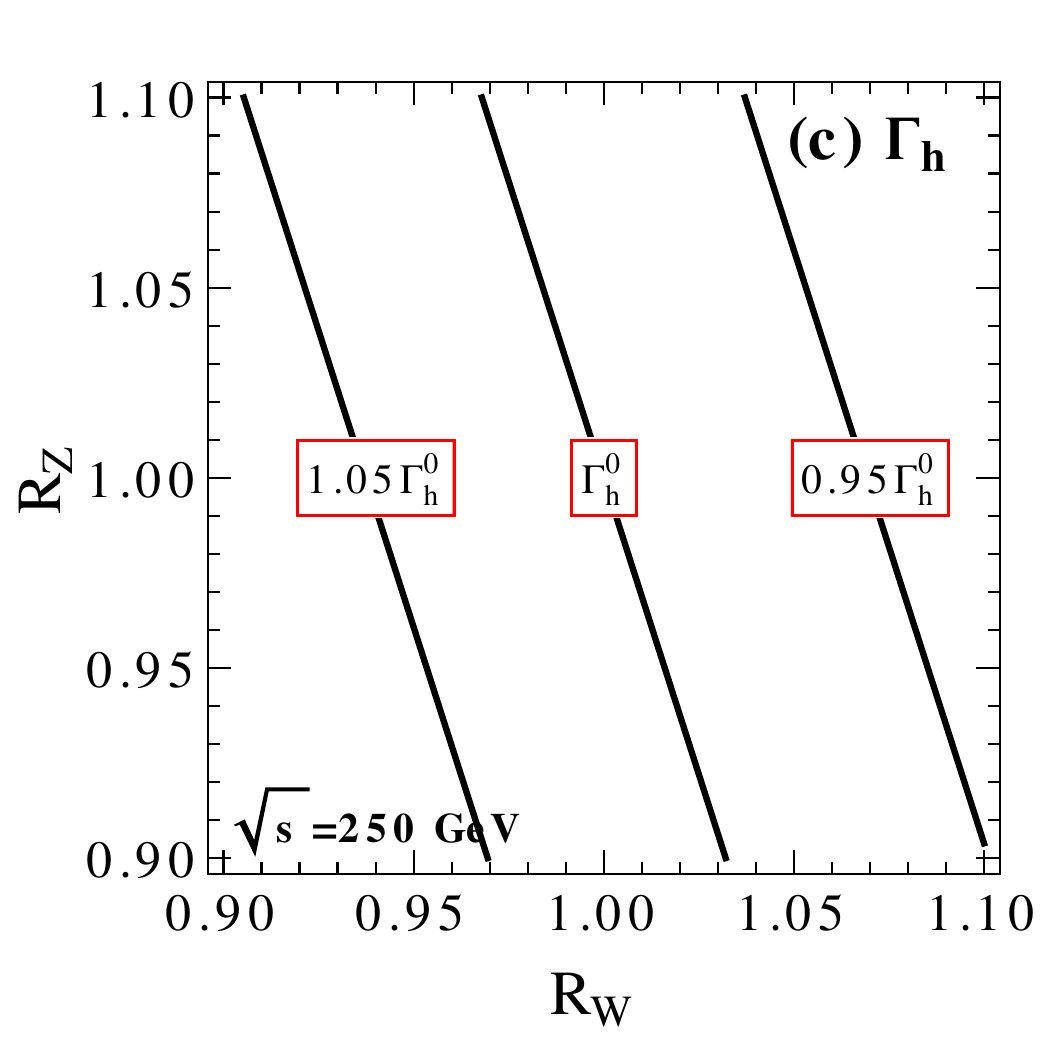}
\includegraphics[scale=0.24]{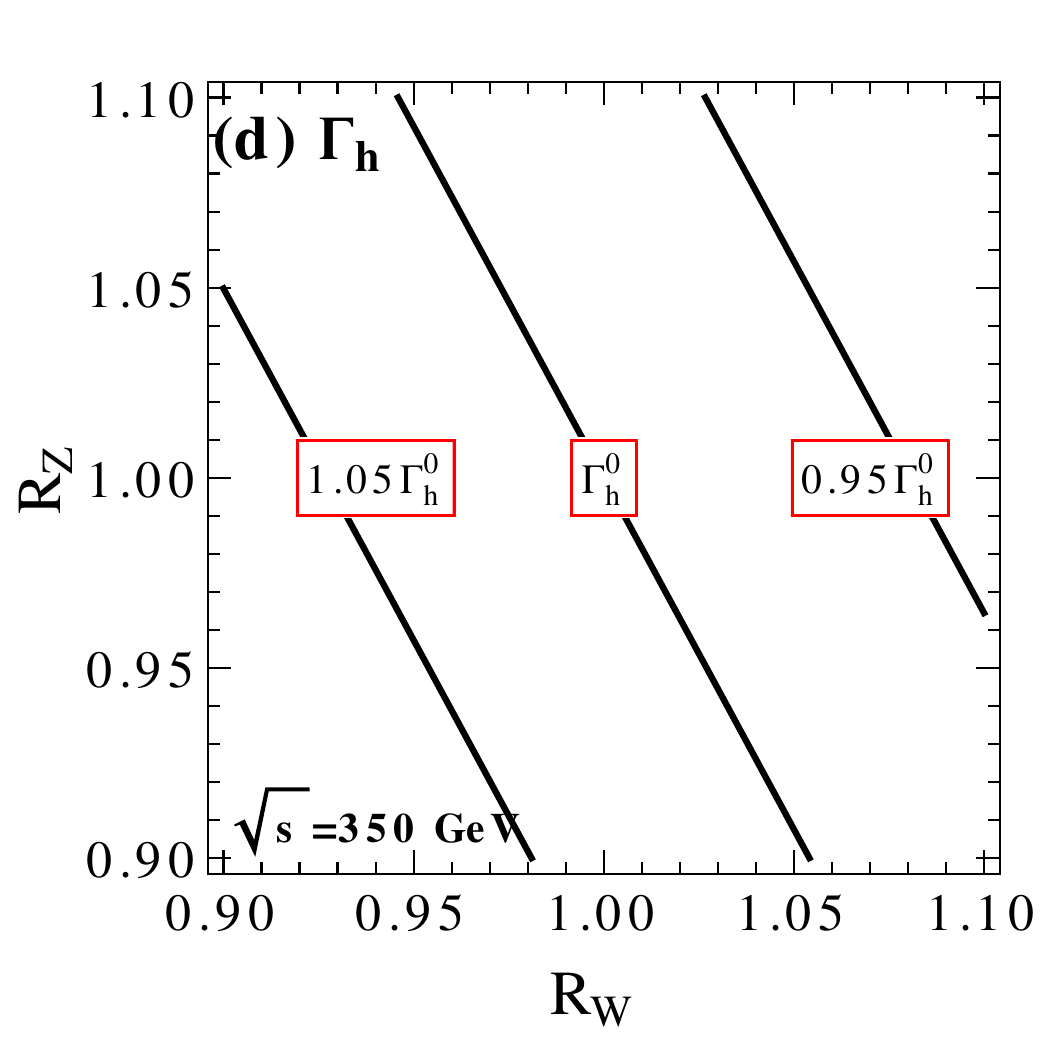}
\caption{The contours of $\Gamma_h$ in the plane of ($R_{cs}$, $R_{Zh}$) (a, b) and ($R_W$,$R_Z$) (c, d)  at $\sqrt{s}=$250 and 350 GeV. $\Gamma_h^0=4.07~{\rm MeV}$ is the Higgs width in the SM. }
\label{fig:Gammh}
\end{center}
\end{figure}

To determine of the Higgs width, we choose two benchmark of collider energies $\sqrt{s}=250~{\rm GeV}$ and $\sqrt{s}=350~{\rm GeV}$ in our study.
The ratios $R_{Zh}$ and $R_{cs}$ are,
\begin{align}
&250~{\rm GeV}: R_{Zh}=1-c_H+2.07c_{BB}+0.81c_{HB}+6.32c_{HW};\nn\\
&250~{\rm GeV}: R_{cs}=1+2.07c_{BB}+0.81c_{HB}+8.25c_{HW};\nn\\
&350~{\rm GeV}: R_{Zh}=1-c_H+2.76c_{BB}+2.54c_{HB}+17.59c_{HW};\nn\\
&350~{\rm GeV}: R_{cs}=1+2.77c_{BB}+2.54c_{HB}+20.48c_{HW}.
\end{align}
The Higgs width is,
\begin{align}
&250~{\rm GeV}: \Gamma_h=\frac{0.174R_{\rm cs}-0.483R_{Zh}-0.174}{{\rm BR}_\gamma-1.712{\rm BR}_W-4.496{\rm BR}_Z}\Gamma_h^0,\nn\\
&350~{\rm GeV}: \Gamma_h=\frac{0.406R_{\rm cs}-0.575R_{Zh}-0.406}{{\rm BR}_\gamma-1.750{\rm BR}_W-7.732{\rm BR}_Z} \Gamma_h^0.
\label{eq:gamma2}
\end{align}
We plot the contours of $\Gamma_h$ in the plane of $R_{cs}$ and $R_{Zh}$ at $\sqrt{s}=$250 and 350 GeV with the SM branching ratios ($\rm {BR}_\gamma^{\rm SM}=2.27\times 10^{-3}$, $\rm {BR}_W^{\rm SM}=0.214$ and $\rm {BR}_Z^{\rm SM}=0.0262$~\cite{Tanabashi:2018oca}. ) in Fig.~\ref{fig:Gammh}(a) and (b).  The Higgs boson width in the SM prediction $\Gamma_h^0=4.07~{\rm MeV}$ is used for reference.  It shows that $\Gamma_h$ is more sensitive to $R_{Zh}$ than $R_{cs}$ at $\sqrt{s}=250~{\rm GeV}$ (see Eq.~\ref{eq:gamma2}). However, with the increase of the collider energy, the cross section of $e^+e^-\to\nu_e\bar{\nu}_eh$ becomes larger, so that a stronger dependence on $R_{cs}$ is found in Fig.~\ref{fig:Gammh}(b) (see Eq.~\ref{eq:gamma2}). The slopes of the contours are depending on the ratio $a/b$ in  Eq.~\ref{eq:width}. We also plot the contours of $\Gamma_h$ in the plane ($R_W$,$R_Z$) at  $\sqrt{s}=$250 and 350 GeV with $\rm {BR}_\gamma=2.27\times 10^{-3}$ and $R_{cs/Zh}=1$.  The slopes are depending on the  ratio $c\rm{BR}_{W}^{\rm SM}/(d \rm {BR}_Z^{\rm SM})$.

\subsection{Error analysis}
Now we discuss the uncertainty of $\Gamma_h$ from the experimental measurements. 
Based on the error propagation equation, we obtain the error of $\Gamma_h$, which is normalized to the central value $\Gamma_h^0$, is
\begin{align}
\left(\dfrac{\delta\Gamma_h}{\Gamma_h^0}\right)^2&=\dfrac{a^2(\delta R_{cs})^2+b^2(\delta R_{Zh})^2}{\left(a(R_{cs}^0-1)+bR_{Zh}^0\right)^2}\nn\\
&+\dfrac{(\delta R_\gamma)^2+c^2(\delta R_W)^2+d^2(\delta R_Z)^2}{\left({\rm BR}_\gamma^{\rm SM}R_\gamma^0+c{\rm BR}_W^{\rm SM}R_W^0+d{\rm BR}_Z^{\rm SM}R_Z^0\right)^2},
\end{align}
where $R_i^0$ and $\delta R_i$ are the central values and errors of those observables, respectively. The uncertainties of $R_i$ are given by,
\begin{align}
&\dfrac{\delta R_{cs}}{R_{cs}^0}=\sqrt{\left(\dfrac{\delta\sigma_{Zh,b}}{\sigma_{Zh,b}^0}\right)^2+\left(\dfrac{\delta\sigma_{\nu\bar{\nu}h,b}}{\sigma_{\nu\bar{\nu}h,b}^0}\right)^2},\nn\\
&\dfrac{\delta R_{Zh}}{R_{Zh}^0}=\frac{\delta \sigma_{Zh}}{\sigma_{Zh}^0},\nn\\
&\dfrac{\delta R_{W,Z,\gamma}}{ R_{W,Z,\gamma}^0}=\sqrt{\left(\dfrac{\delta\sigma_{Zh,W/Z/\gamma}}{\sigma_{Zh,W/Z/\gamma}^0}\right)^2+\left(\dfrac{\delta\sigma_{Zh}}{\sigma_{Zh}^0}\right)^2},
\end{align}
where $\sigma_{Zh,i}^0$ ($\delta \sigma_{Zh,i}$) and  $\sigma_{\nu\bar{\nu}h,b}^0$ ($\delta \sigma_{\nu\bar{\nu}h,b}$) are the central values (errrors) of the processes $e^+e^-\to Zh(\to ii)$ and $e^+e^-\to \nu_e\bar{\nu}_e h(\to b\bar{b})$, respectively, and  $\sigma_{Zh}^0$ ($\delta \sigma_{Zh}$) is the central value (error) of the inclusive cross section of the process $e^+e^-\to Zh$. 

The expected uncertainties of the cross sections at the CEPC with $\sqrt{s}=250~{\rm GeV}$ and an integrated luminosity ($\mathcal{L}$) of $5~{\rm ab}^{-1}$ are~\cite{CDR,CEPCStudyGroup:2018ghi},
\begin{align}
&\frac{\delta\sigma_{Zh}}{\sigma_{Zh}^0}=0.51\%,\quad \dfrac{\delta \sigma_{Zh,b}}{\sigma_{Zh,b}^0}=0.28\%,\quad \dfrac{\delta\sigma_{\nu\bar{\nu}h,b}}{\sigma_{\nu\bar{\nu}h,b}^0}=2.8\%\nn\\
&\dfrac{\delta\sigma_{Zh,Z}}{\sigma_{Zh,Z}^0}=4.3\%,\quad \dfrac{\delta\sigma_{Zh,W}}{\sigma_{Zh,W}^0}=1.5\%,\quad \dfrac{\delta\sigma_{Zh,\gamma}}{\sigma_{Zh,\gamma}^0}=9\%.
\end{align}
Therefore,  we obtain the uncertainties of the $R_i$,
\begin{align}
&\dfrac{\delta R_{cs}}{R_{cs}^0}=2.8\%,& &\dfrac{\delta R_{Zh}}{R_{Zh}^0}=0.51\%,& \dfrac{\delta R_Z}{R_Z^0}=4.3\%,\nn\\
& \dfrac{\delta R_W}{R_W^0}=1.6\%, &&\dfrac{\delta R_\gamma}{R_\gamma^0}=9\%.
\end{align}
In the following analysis, we will assume the central values of $R_i$ to be same with SM predictions, i.e. $R_i^0=1$.  As a result, the error of $\Gamma_h$ at $\sqrt{s}=250~{\rm GeV}$ and $\mathcal{L}=5~{\rm ab}^{-1}$ is,
\begin{align}
\left(\dfrac{\delta\Gamma_h}{\Gamma_h^0}\right)^2&=0.13\left(\delta R_{cs}\right)^2+\left(\delta R_{Zh}\right)^2+2.2\times 10^{-5}\left(\delta R_\gamma \right)^2\nn\\
&+0.58\left(\delta R_W\right)^2+0.059\left(\delta R_Z\right)^2.
\end{align}
Clearly, the cross section ratio $R_{cs}$ and branching ratios ${\rm BR}_{W/Z}$ dominant the uncertainty of the Higgs width. 
For the general collider energy and luminosity, we could rescale the relative errors by the following method,
\bea
\dfrac{\delta\sigma_A^i}{\delta\sigma_B^i}=\sqrt{\dfrac{\sigma_B^i(s_B)\mathcal{L}_B}{\sigma_A^i(s_A)\mathcal{L}_A}},
\eea
where $\delta\sigma_{A,B}^i$ and $\sigma_{A,B}^i(s_{A,B})$ are  the cross section error and  central value of process  $i$  with collider energy $s_{A,B}$ and  integrated luminosity $\mathcal{L}_{A,B}$, respectively.

\begin{figure}
\begin{center}
\includegraphics[scale=0.33]{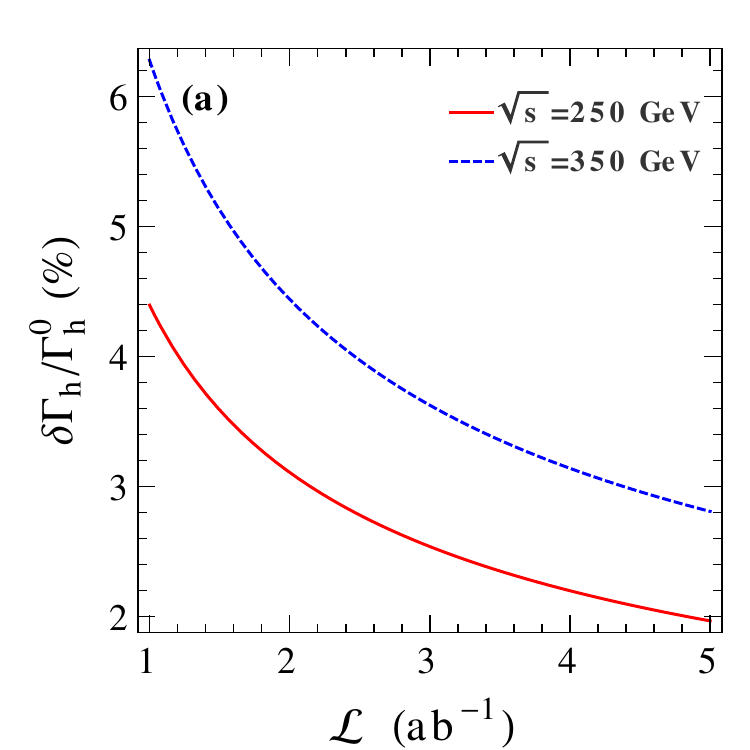}
\includegraphics[scale=0.234]{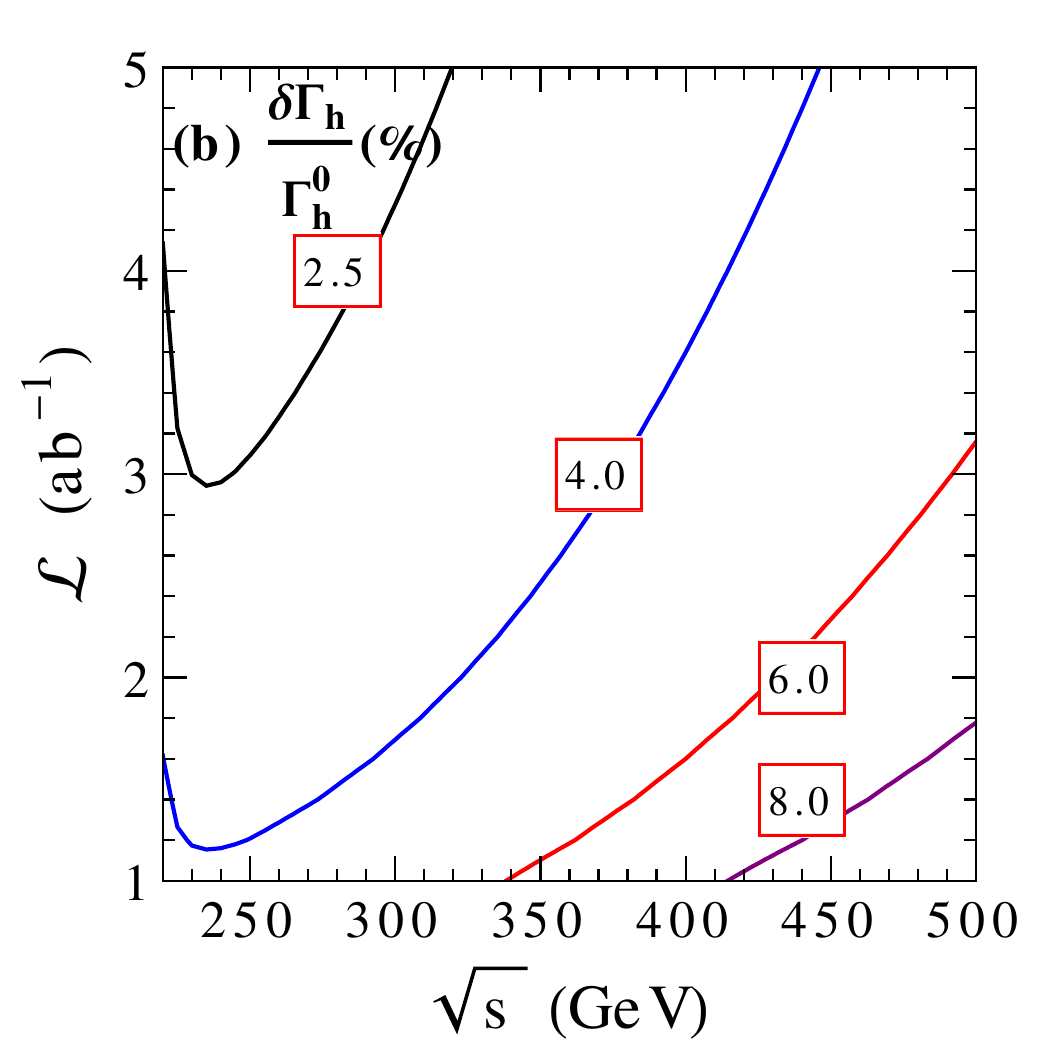}
\caption{(a) The uncertainty of $\Gamma_h$ with the integrated luminosity $\mathcal{L} (\rm{ab}^{-1})$ at $\sqrt{s}=250~{\rm GeV}$ and $\sqrt{s}=350~{\rm GeV}$; (b) The contours of uncertainty of $\delta\Gamma_h$ in the plane of the collider energy $\sqrt{s}$ (GeV) and integrated luminosity $\mathcal{L}~(\rm ab^{-1}$).}
\label{fig:errgam}
\end{center}
\end{figure} 
Figure~\ref{fig:errgam} (a) displays the relative error of $\Gamma_h$ with the integrated luminosity $\mathcal{L}~(\rm{ab}^{-1})$ at $\sqrt{s}=250~{\rm GeV}$ (red solid line) and $\sqrt{s}=350~{\rm GeV}$ (blue dashed line). It shows that $\Gamma_h$ could be measured with an accuracy of percentage; e.g. $\delta\Gamma_h/\Gamma_h^0=2.0\%$ at $\sqrt{s}=250~{\rm GeV}$ and  $\delta\Gamma_h/\Gamma_h^0=2.8\%$ at $\sqrt{s}=350~{\rm GeV}$ with $\mathcal{L}=5~{\rm ab}^{-1}$. Figure~\ref{fig:errgam} (b) shows the contours of $\delta\Gamma_h/\Gamma_h^0$ in the plane of the collider energy $\sqrt{s}$ (GeV) and integrated luminosity $\mathcal{L}~({\rm ab}^{-1})$.

\section{Limiting fermion Yukawa couplings}
\label{sec:NP}
In this section, we combine the branching ratio ${\rm BR}(h\to f\bar{f})$ and $\Gamma_h$ measurements to constrain the fermion Yukawa couplings.  The effective Lagrangian of the $hf\bar{f}$ interaction could be parametrized by,
\beq
\mathcal{L}_{Yukawa}=-\frac{m_f}{v}\kappa_f\bar{f}fh,
\eeq
where $m_f$ is the mass of fermion $f=b,c,\tau,\mu$ and $\kappa_f=1$ in the SM. The $\kappa_f$ could by obtained through the Higgs decay branching ratio and $\Gamma_h$ measurements, i.e.
\beq
{\rm BR}(h\to f\bar{f})=\frac{\kappa_f^2\Gamma_{f\bar{f}}^{\rm SM}}{\Gamma_h}=\frac{\sigma_{Zh,f}}{\sigma_{Zh}},
\eeq
where $\Gamma_{f\bar{f}}^{\rm SM}$ is the partial decay width of $h\to f\bar{f}$ in the SM.
Therefore, we have
\beq
\kappa_f=\sqrt{\frac{\sigma_{Zh,f}}{\sigma_{Zh}}\frac{\Gamma_h}{\Gamma_{f\bar{f}}^{\rm SM}}}.
\eeq
The uncertainty of $\kappa_f$ is,
\beq
\left(\frac{\delta \kappa_f}{\kappa_f^0}\right)^2=\frac{1}{4}\left[\left(\frac{\delta \sigma_{Zh,f}}{\sigma_{Zh,f}^0}\right)^2+\left(\frac{\delta \sigma_{Zh}}{\sigma_{Zh}^0}\right)^2+\left(\frac{\delta \Gamma_h}{\Gamma_h^0}\right)^2\right],
\eeq
where $\kappa_f^0$ is the central value of the $\kappa_f$. The expected uncertainties of the cross sections at  $\sqrt{s}=250~{\rm GeV}$ with $\mathcal{L}=5~{\rm ab}^{-1}$ are~\cite{CDR,CEPCStudyGroup:2018ghi},
\begin{align}
&\dfrac{\delta \sigma_{Zh,b}}{\sigma_{Zh,b}^0}=0.28\%,& &\dfrac{\delta \sigma_{Zh,c}}{\sigma_{Zh,c}^0}=3.3\%,\nn\\
&\dfrac{\delta \sigma_{Zh,\tau}}{\sigma_{Zh,\tau}^0}=0.8\%,&&\dfrac{\delta \sigma_{Zh,\mu}}{\sigma_{Zh,\mu}^0}=17\%.
\end{align}
That yields a error on the fermion Yukawa couplings as
\begin{align}
&\frac{\delta \kappa_b}{\kappa_b^0}=1.04\%, && \frac{\delta \kappa_c}{\kappa_c^0}=1.95\%, &&\frac{\delta \kappa_\tau}{\kappa_\tau^0}=1.11\%, &
& \frac{\delta \kappa_\mu}{\kappa_\mu^0}=8.56\%.
\end{align}
We emphasize that the limits for the fermion Yukawa couplings are totally model-independent.

\section{Conclusions}
\label{sec:con}
In this work we proposed a method to probe the Higgs width within the model-independent framework of the Standard Model Effective Field Theory  at the future $e^+e^-$ collider. The effects of the new physics are parameterized by a set of the dimension-6 operators in the SMEFT.  We compute the Higgs production cross sections and decay branching ratios from the contribution of  BSM operators. It shows that the Higgs width could be determined after we combine the cross sections of  $e^+e^-\to Zh$ and $e^+e^-\to \nu_e\bar{\nu}_e h$ production processes and the branching ratio measurements $h\to WW^*/ZZ^*/\gamma\gamma$.  We note that the size of the Higgs width is not sensitive to the ${\rm BR}(h\to\gamma\gamma)$ and its impact can be ignored during the numerical calculation. We further demonstrate  that the Higgs width could be constrained to be percentage level at $\sqrt{s}=250$ and 350 GeV with integrated luminosity $5~{\rm ab}^{-1}$.   As an application, we combine the Higgs width information and the decay branching ratios to constrain the fermion Yukawa couplings.

\begin{acknowledgments}
The author thank Ling-Xiao Xu and Zhite Yu for the collaboration at the early stage of the project, and Y. Liu, C.-P. Yuan for helpful discussions and comments. This work is supported by the U.S. Department of Energy, Office of Science, Office of Nuclear Physics, under Contract DE-AC52- 06NA25396 through the LANL/LDRD Program.
\end{acknowledgments}
\bibliographystyle{apsrev}
\bibliography{reference}

\begin{thebibliography}{42}
\expandafter\ifx\csname natexlab\endcsname\relax\def\natexlab#1{#1}\fi
\expandafter\ifx\csname bibnamefont\endcsname\relax
  \def\bibnamefont#1{#1}\fi
\expandafter\ifx\csname bibfnamefont\endcsname\relax
  \def\bibfnamefont#1{#1}\fi
\expandafter\ifx\csname citenamefont\endcsname\relax
  \def\citenamefont#1{#1}\fi
\expandafter\ifx\csname url\endcsname\relax
  \def\url#1{\texttt{#1}}\fi
\expandafter\ifx\csname urlprefix\endcsname\relax\def\urlprefix{URL }\fi
\providecommand{\bibinfo}[2]{#2}
\providecommand{\eprint}[2][]{\url{#2}}

\bibitem[{\citenamefont{Caola and Melnikov}(2013)}]{Caola:2013yja}
\bibinfo{author}{\bibfnamefont{F.}~\bibnamefont{Caola}} \bibnamefont{and}
  \bibinfo{author}{\bibfnamefont{K.}~\bibnamefont{Melnikov}},
  \bibinfo{journal}{Phys. Rev.} \textbf{\bibinfo{volume}{D88}},
  \bibinfo{pages}{054024} (\bibinfo{year}{2013}), \eprint{1307.4935}.

\bibitem[{\citenamefont{Campbell
  et~al.}(2014{\natexlab{a}})\citenamefont{Campbell, Ellis, and
  Williams}}]{Campbell:2013una}
\bibinfo{author}{\bibfnamefont{J.~M.} \bibnamefont{Campbell}},
  \bibinfo{author}{\bibfnamefont{R.~K.} \bibnamefont{Ellis}}, \bibnamefont{and}
  \bibinfo{author}{\bibfnamefont{C.}~\bibnamefont{Williams}},
  \bibinfo{journal}{JHEP} \textbf{\bibinfo{volume}{04}}, \bibinfo{pages}{060}
  (\bibinfo{year}{2014}{\natexlab{a}}), \eprint{1311.3589}.

\bibitem[{\citenamefont{Campbell
  et~al.}(2014{\natexlab{b}})\citenamefont{Campbell, Ellis, and
  Williams}}]{Campbell:2013wga}
\bibinfo{author}{\bibfnamefont{J.~M.} \bibnamefont{Campbell}},
  \bibinfo{author}{\bibfnamefont{R.~K.} \bibnamefont{Ellis}}, \bibnamefont{and}
  \bibinfo{author}{\bibfnamefont{C.}~\bibnamefont{Williams}},
  \bibinfo{journal}{Phys. Rev.} \textbf{\bibinfo{volume}{D89}},
  \bibinfo{pages}{053011} (\bibinfo{year}{2014}{\natexlab{b}}),
  \eprint{1312.1628}.

\bibitem[{\citenamefont{Dixon and Li}(2013)}]{Dixon:2013haa}
\bibinfo{author}{\bibfnamefont{L.~J.} \bibnamefont{Dixon}} \bibnamefont{and}
  \bibinfo{author}{\bibfnamefont{Y.}~\bibnamefont{Li}}, \bibinfo{journal}{Phys.
  Rev. Lett.} \textbf{\bibinfo{volume}{111}}, \bibinfo{pages}{111802}
  (\bibinfo{year}{2013}), \eprint{1305.3854}.

\bibitem[{\citenamefont{Campbell et~al.}(2017)\citenamefont{Campbell, Carena,
  Harnik, and Liu}}]{Campbell:2017rke}
\bibinfo{author}{\bibfnamefont{J.}~\bibnamefont{Campbell}},
  \bibinfo{author}{\bibfnamefont{M.}~\bibnamefont{Carena}},
  \bibinfo{author}{\bibfnamefont{R.}~\bibnamefont{Harnik}}, \bibnamefont{and}
  \bibinfo{author}{\bibfnamefont{Z.}~\bibnamefont{Liu}},
  \bibinfo{journal}{Phys. Rev. Lett.} \textbf{\bibinfo{volume}{119}},
  \bibinfo{pages}{181801} (\bibinfo{year}{2017}), \bibinfo{note}{[Addendum:
  Phys. Rev. Lett.119,no.19,199901(2017)]}, \eprint{1704.08259}.

\bibitem[{\citenamefont{Cao et~al.}(2017)\citenamefont{Cao, Chen, and
  Liu}}]{Cao:2016wib}
\bibinfo{author}{\bibfnamefont{Q.-H.} \bibnamefont{Cao}},
  \bibinfo{author}{\bibfnamefont{S.-L.} \bibnamefont{Chen}}, \bibnamefont{and}
  \bibinfo{author}{\bibfnamefont{Y.}~\bibnamefont{Liu}},
  \bibinfo{journal}{Phys. Rev.} \textbf{\bibinfo{volume}{D95}},
  \bibinfo{pages}{053004} (\bibinfo{year}{2017}), \eprint{1602.01934}.

\bibitem[{\citenamefont{Cao et~al.}(2019{\natexlab{a}})\citenamefont{Cao, Chen,
  Liu, Zhang, and Zhang}}]{Cao:2019ygh}
\bibinfo{author}{\bibfnamefont{Q.-H.} \bibnamefont{Cao}},
  \bibinfo{author}{\bibfnamefont{S.-L.} \bibnamefont{Chen}},
  \bibinfo{author}{\bibfnamefont{Y.}~\bibnamefont{Liu}},
  \bibinfo{author}{\bibfnamefont{R.}~\bibnamefont{Zhang}}, \bibnamefont{and}
  \bibinfo{author}{\bibfnamefont{Y.}~\bibnamefont{Zhang}},
  \bibinfo{journal}{Phys. Rev. D} \textbf{\bibinfo{volume}{99}},
  \bibinfo{pages}{113003} (\bibinfo{year}{2019}{\natexlab{a}}),
  \eprint{1901.04567}.

\bibitem[{\citenamefont{Aad et~al.}(2015)}]{Aad:2015xua}
\bibinfo{author}{\bibfnamefont{G.}~\bibnamefont{Aad}} \bibnamefont{et~al.}
  (\bibinfo{collaboration}{ATLAS}), \bibinfo{journal}{Eur. Phys. J.}
  \textbf{\bibinfo{volume}{C75}}, \bibinfo{pages}{335} (\bibinfo{year}{2015}),
  \eprint{1503.01060}.

\bibitem[{\citenamefont{Aaboud et~al.}(2018)}]{Aaboud:2018puo}
\bibinfo{author}{\bibfnamefont{M.}~\bibnamefont{Aaboud}} \bibnamefont{et~al.}
  (\bibinfo{collaboration}{ATLAS}) (\bibinfo{year}{2018}), \eprint{1808.01191}.

\bibitem[{\citenamefont{Khachatryan et~al.}(2014)}]{Khachatryan:2014iha}
\bibinfo{author}{\bibfnamefont{V.}~\bibnamefont{Khachatryan}}
  \bibnamefont{et~al.} (\bibinfo{collaboration}{CMS}), \bibinfo{journal}{Phys.
  Lett.} \textbf{\bibinfo{volume}{B736}}, \bibinfo{pages}{64}
  (\bibinfo{year}{2014}), \eprint{1405.3455}.

\bibitem[{\citenamefont{Khachatryan et~al.}(2016)}]{Khachatryan:2016ctc}
\bibinfo{author}{\bibfnamefont{V.}~\bibnamefont{Khachatryan}}
  \bibnamefont{et~al.} (\bibinfo{collaboration}{CMS}), \bibinfo{journal}{JHEP}
  \textbf{\bibinfo{volume}{09}}, \bibinfo{pages}{051} (\bibinfo{year}{2016}),
  \eprint{1605.02329}.

\bibitem[{\citenamefont{Collaboration}(2018)}]{CMS:2018bwq}
\bibinfo{author}{\bibfnamefont{C.}~\bibnamefont{Collaboration}}
  (\bibinfo{collaboration}{CMS}) (\bibinfo{year}{2018}).

\bibitem[{CEP(2018)}]{CEPCStudyGroup:2018rmc}
 (\bibinfo{year}{2018}), \eprint{1809.00285}.

\bibitem[{\citenamefont{Bicer et~al.}(2014)}]{Gomez-Ceballos:2013zzn}
\bibinfo{author}{\bibfnamefont{M.}~\bibnamefont{Bicer}} \bibnamefont{et~al.}
  (\bibinfo{collaboration}{TLEP Design Study Working Group}),
  \bibinfo{journal}{JHEP} \textbf{\bibinfo{volume}{01}}, \bibinfo{pages}{164}
  (\bibinfo{year}{2014}), \eprint{1308.6176}.

\bibitem[{\citenamefont{Baer et~al.}(2013)\citenamefont{Baer, Barklow, Fujii,
  Gao, Hoang, Kanemura, List, Logan, Nomerotski, Perelstein
  et~al.}}]{Baer:2013cma}
\bibinfo{author}{\bibfnamefont{H.}~\bibnamefont{Baer}},
  \bibinfo{author}{\bibfnamefont{T.}~\bibnamefont{Barklow}},
  \bibinfo{author}{\bibfnamefont{K.}~\bibnamefont{Fujii}},
  \bibinfo{author}{\bibfnamefont{Y.}~\bibnamefont{Gao}},
  \bibinfo{author}{\bibfnamefont{A.}~\bibnamefont{Hoang}},
  \bibinfo{author}{\bibfnamefont{S.}~\bibnamefont{Kanemura}},
  \bibinfo{author}{\bibfnamefont{J.}~\bibnamefont{List}},
  \bibinfo{author}{\bibfnamefont{H.~E.} \bibnamefont{Logan}},
  \bibinfo{author}{\bibfnamefont{A.}~\bibnamefont{Nomerotski}},
  \bibinfo{author}{\bibfnamefont{M.}~\bibnamefont{Perelstein}},
  \bibnamefont{et~al.} (\bibinfo{year}{2013}), \eprint{1306.6352}.

\bibitem[{\citenamefont{Han et~al.}(2014)\citenamefont{Han, Liu, and
  Sayre}}]{Han:2013kya}
\bibinfo{author}{\bibfnamefont{T.}~\bibnamefont{Han}},
  \bibinfo{author}{\bibfnamefont{Z.}~\bibnamefont{Liu}}, \bibnamefont{and}
  \bibinfo{author}{\bibfnamefont{J.}~\bibnamefont{Sayre}},
  \bibinfo{journal}{Phys. Rev.} \textbf{\bibinfo{volume}{D89}},
  \bibinfo{pages}{113006} (\bibinfo{year}{2014}), \eprint{1311.7155}.

\bibitem[{\citenamefont{Asner et~al.}(2013)}]{Asner:2013psa}
\bibinfo{author}{\bibfnamefont{D.~M.} \bibnamefont{Asner}}
  \bibnamefont{et~al.}, in \emph{\bibinfo{booktitle}{{Proceedings, 2013
  Community Summer Study on the Future of U.S. Particle Physics: Snowmass on
  the Mississippi (CSS2013): Minneapolis, MN, USA, July 29-August 6, 2013}}}
  (\bibinfo{year}{2013}), \eprint{1310.0763},
  \urlprefix\url{http://www.slac.stanford.edu/econf/C1307292/docs/submittedArxivFiles/1310.0763.pdf}.

\bibitem[{\citenamefont{Ahmad et~al.}(2015)}]{CDR}
\bibinfo{author}{\bibfnamefont{M.}~\bibnamefont{Ahmad}} \bibnamefont{et~al.}
  (\bibinfo{collaboration}{CEPC-SPPC Study Group}),
  \emph{\bibinfo{title}{{CEPC-SPPC Preliminary Conceptual Design Report. 1.
  Physics and Detector}}} (\bibinfo{year}{2015}),
  \urlprefix\url{http://cepc.ihep.ac.cn/preCDR/main_preCDR.pdf}.

\bibitem[{\citenamefont{Dürig et~al.}(2014)\citenamefont{Dürig, Fujii, List,
  and Tian}}]{Durig:2014lfa}
\bibinfo{author}{\bibfnamefont{C.}~\bibnamefont{Dürig}},
  \bibinfo{author}{\bibfnamefont{K.}~\bibnamefont{Fujii}},
  \bibinfo{author}{\bibfnamefont{J.}~\bibnamefont{List}}, \bibnamefont{and}
  \bibinfo{author}{\bibfnamefont{J.}~\bibnamefont{Tian}}, in
  \emph{\bibinfo{booktitle}{{International Workshop on Future Linear Colliders
  (LCWS13) Tokyo, Japan, November 11-15, 2013}}} (\bibinfo{year}{2014}),
  \eprint{1403.7734}.

\bibitem[{\citenamefont{Chen and Ruan}(2016)}]{Chen:2016prx}
\bibinfo{author}{\bibfnamefont{Z.}~\bibnamefont{Chen}} \bibnamefont{and}
  \bibinfo{author}{\bibfnamefont{M.}~\bibnamefont{Ruan}}
  (\bibinfo{collaboration}{CEPC}), \bibinfo{journal}{PoS}
  \textbf{\bibinfo{volume}{ICHEP2016}}, \bibinfo{pages}{432}
  (\bibinfo{year}{2016}).

\bibitem[{\citenamefont{Barklow et~al.}(2018)\citenamefont{Barklow, Fujii,
  Jung, Karl, List, Ogawa, Peskin, and Tian}}]{Barklow:2017suo}
\bibinfo{author}{\bibfnamefont{T.}~\bibnamefont{Barklow}},
  \bibinfo{author}{\bibfnamefont{K.}~\bibnamefont{Fujii}},
  \bibinfo{author}{\bibfnamefont{S.}~\bibnamefont{Jung}},
  \bibinfo{author}{\bibfnamefont{R.}~\bibnamefont{Karl}},
  \bibinfo{author}{\bibfnamefont{J.}~\bibnamefont{List}},
  \bibinfo{author}{\bibfnamefont{T.}~\bibnamefont{Ogawa}},
  \bibinfo{author}{\bibfnamefont{M.~E.} \bibnamefont{Peskin}},
  \bibnamefont{and} \bibinfo{author}{\bibfnamefont{J.}~\bibnamefont{Tian}},
  \bibinfo{journal}{Phys. Rev.} \textbf{\bibinfo{volume}{D97}},
  \bibinfo{pages}{053003} (\bibinfo{year}{2018}), \eprint{1708.08912}.

\bibitem[{\citenamefont{Lafaye et~al.}(2017)\citenamefont{Lafaye, Plehn, Rauch,
  and Zerwas}}]{Lafaye:2017kgf}
\bibinfo{author}{\bibfnamefont{R.}~\bibnamefont{Lafaye}},
  \bibinfo{author}{\bibfnamefont{T.}~\bibnamefont{Plehn}},
  \bibinfo{author}{\bibfnamefont{M.}~\bibnamefont{Rauch}}, \bibnamefont{and}
  \bibinfo{author}{\bibfnamefont{D.}~\bibnamefont{Zerwas}},
  \bibinfo{journal}{Phys. Rev.} \textbf{\bibinfo{volume}{D96}},
  \bibinfo{pages}{075044} (\bibinfo{year}{2017}), \eprint{1706.02174}.

\bibitem[{\citenamefont{Buchmuller and Wyler}(1986)}]{Buchmuller:1985jz}
\bibinfo{author}{\bibfnamefont{W.}~\bibnamefont{Buchmuller}} \bibnamefont{and}
  \bibinfo{author}{\bibfnamefont{D.}~\bibnamefont{Wyler}},
  \bibinfo{journal}{Nucl. Phys. B} \textbf{\bibinfo{volume}{268}},
  \bibinfo{pages}{621} (\bibinfo{year}{1986}).

\bibitem[{\citenamefont{Giudice et~al.}(2007)\citenamefont{Giudice, Grojean,
  Pomarol, and Rattazzi}}]{Giudice:2007fh}
\bibinfo{author}{\bibfnamefont{G.~F.} \bibnamefont{Giudice}},
  \bibinfo{author}{\bibfnamefont{C.}~\bibnamefont{Grojean}},
  \bibinfo{author}{\bibfnamefont{A.}~\bibnamefont{Pomarol}}, \bibnamefont{and}
  \bibinfo{author}{\bibfnamefont{R.}~\bibnamefont{Rattazzi}},
  \bibinfo{journal}{JHEP} \textbf{\bibinfo{volume}{06}}, \bibinfo{pages}{045}
  (\bibinfo{year}{2007}), \eprint{hep-ph/0703164}.

\bibitem[{\citenamefont{Grzadkowski et~al.}(2010)\citenamefont{Grzadkowski,
  Iskrzynski, Misiak, and Rosiek}}]{Grzadkowski:2010es}
\bibinfo{author}{\bibfnamefont{B.}~\bibnamefont{Grzadkowski}},
  \bibinfo{author}{\bibfnamefont{M.}~\bibnamefont{Iskrzynski}},
  \bibinfo{author}{\bibfnamefont{M.}~\bibnamefont{Misiak}}, \bibnamefont{and}
  \bibinfo{author}{\bibfnamefont{J.}~\bibnamefont{Rosiek}},
  \bibinfo{journal}{JHEP} \textbf{\bibinfo{volume}{10}}, \bibinfo{pages}{085}
  (\bibinfo{year}{2010}), \eprint{1008.4884}.

\bibitem[{\citenamefont{Li et~al.}(2020)\citenamefont{Li, Ren, Shu, Xiao, Yu,
  and Zheng}}]{Li:2020gnx}
\bibinfo{author}{\bibfnamefont{H.-L.} \bibnamefont{Li}},
  \bibinfo{author}{\bibfnamefont{Z.}~\bibnamefont{Ren}},
  \bibinfo{author}{\bibfnamefont{J.}~\bibnamefont{Shu}},
  \bibinfo{author}{\bibfnamefont{M.-L.} \bibnamefont{Xiao}},
  \bibinfo{author}{\bibfnamefont{J.-H.} \bibnamefont{Yu}}, \bibnamefont{and}
  \bibinfo{author}{\bibfnamefont{Y.-H.} \bibnamefont{Zheng}}
  (\bibinfo{year}{2020}), \eprint{2005.00008}.

\bibitem[{\citenamefont{De~Blas et~al.}(2019)\citenamefont{De~Blas, Durieux,
  Grojean, Gu, and Paul}}]{DeBlas:2019qco}
\bibinfo{author}{\bibfnamefont{J.}~\bibnamefont{De~Blas}},
  \bibinfo{author}{\bibfnamefont{G.}~\bibnamefont{Durieux}},
  \bibinfo{author}{\bibfnamefont{C.}~\bibnamefont{Grojean}},
  \bibinfo{author}{\bibfnamefont{J.}~\bibnamefont{Gu}}, \bibnamefont{and}
  \bibinfo{author}{\bibfnamefont{A.}~\bibnamefont{Paul}},
  \bibinfo{journal}{JHEP} \textbf{\bibinfo{volume}{12}}, \bibinfo{pages}{117}
  (\bibinfo{year}{2019}), \eprint{1907.04311}.

\bibitem[{\citenamefont{Ge et~al.}(2016)\citenamefont{Ge, He, and
  Xiao}}]{Ge:2016zro}
\bibinfo{author}{\bibfnamefont{S.-F.} \bibnamefont{Ge}},
  \bibinfo{author}{\bibfnamefont{H.-J.} \bibnamefont{He}}, \bibnamefont{and}
  \bibinfo{author}{\bibfnamefont{R.-Q.} \bibnamefont{Xiao}},
  \bibinfo{journal}{JHEP} \textbf{\bibinfo{volume}{10}}, \bibinfo{pages}{007}
  (\bibinfo{year}{2016}), \eprint{1603.03385}.

\bibitem[{\citenamefont{Chiu et~al.}(2018)\citenamefont{Chiu, Leung, Liu, Lyu,
  and Wang}}]{Chiu:2017yrx}
\bibinfo{author}{\bibfnamefont{W.~H.} \bibnamefont{Chiu}},
  \bibinfo{author}{\bibfnamefont{S.~C.} \bibnamefont{Leung}},
  \bibinfo{author}{\bibfnamefont{T.}~\bibnamefont{Liu}},
  \bibinfo{author}{\bibfnamefont{K.-F.} \bibnamefont{Lyu}}, \bibnamefont{and}
  \bibinfo{author}{\bibfnamefont{L.-T.} \bibnamefont{Wang}},
  \bibinfo{journal}{JHEP} \textbf{\bibinfo{volume}{05}}, \bibinfo{pages}{081}
  (\bibinfo{year}{2018}), \eprint{1711.04046}.

\bibitem[{\citenamefont{Khanpour and
  Mohammadi~Najafabadi}(2017)}]{Khanpour:2017cfq}
\bibinfo{author}{\bibfnamefont{H.}~\bibnamefont{Khanpour}} \bibnamefont{and}
  \bibinfo{author}{\bibfnamefont{M.}~\bibnamefont{Mohammadi~Najafabadi}},
  \bibinfo{journal}{Phys. Rev. D} \textbf{\bibinfo{volume}{95}},
  \bibinfo{pages}{055026} (\bibinfo{year}{2017}), \eprint{1702.00951}.

\bibitem[{\citenamefont{Durieux et~al.}(2017)\citenamefont{Durieux, Grojean,
  Gu, and Wang}}]{Durieux:2017rsg}
\bibinfo{author}{\bibfnamefont{G.}~\bibnamefont{Durieux}},
  \bibinfo{author}{\bibfnamefont{C.}~\bibnamefont{Grojean}},
  \bibinfo{author}{\bibfnamefont{J.}~\bibnamefont{Gu}}, \bibnamefont{and}
  \bibinfo{author}{\bibfnamefont{K.}~\bibnamefont{Wang}},
  \bibinfo{journal}{JHEP} \textbf{\bibinfo{volume}{09}}, \bibinfo{pages}{014}
  (\bibinfo{year}{2017}), \eprint{1704.02333}.

\bibitem[{\citenamefont{Cao et~al.}(2019{\natexlab{b}})\citenamefont{Cao, Xu,
  Yan, and Zhu}}]{Cao:2018cms}
\bibinfo{author}{\bibfnamefont{Q.-H.} \bibnamefont{Cao}},
  \bibinfo{author}{\bibfnamefont{L.-X.} \bibnamefont{Xu}},
  \bibinfo{author}{\bibfnamefont{B.}~\bibnamefont{Yan}}, \bibnamefont{and}
  \bibinfo{author}{\bibfnamefont{S.-H.} \bibnamefont{Zhu}},
  \bibinfo{journal}{Phys. Lett. B} \textbf{\bibinfo{volume}{789}},
  \bibinfo{pages}{233} (\bibinfo{year}{2019}{\natexlab{b}}),
  \eprint{1810.07661}.

\bibitem[{\citenamefont{Xie and Yan}(2021)}]{Xie:2021xtl}
\bibinfo{author}{\bibfnamefont{K.-P.} \bibnamefont{Xie}} \bibnamefont{and}
  \bibinfo{author}{\bibfnamefont{B.}~\bibnamefont{Yan}} (\bibinfo{year}{2021}),
  \eprint{2104.12689}.

\bibitem[{\citenamefont{Dong et~al.}(2018)}]{CEPCStudyGroup:2018ghi}
\bibinfo{author}{\bibfnamefont{M.}~\bibnamefont{Dong}} \bibnamefont{et~al.}
  (\bibinfo{collaboration}{CEPC Study Group}) (\bibinfo{year}{2018}),
  \eprint{1811.10545}.

\bibitem[{Note1()}]{Note1}
Note1, \bibinfo{note}{the Higgs width under the general BSM operators could be
  found in Ref.~\cite {Brivio:2019myy}}.

\bibitem[{\citenamefont{Rizzo}(1980)}]{Rizzo:1980gz}
\bibinfo{author}{\bibfnamefont{T.~G.} \bibnamefont{Rizzo}},
  \bibinfo{journal}{Phys. Rev.} \textbf{\bibinfo{volume}{D22}},
  \bibinfo{pages}{722} (\bibinfo{year}{1980}).

\bibitem[{\citenamefont{Keung and Marciano}(1984)}]{Keung:1984hn}
\bibinfo{author}{\bibfnamefont{W.-Y.} \bibnamefont{Keung}} \bibnamefont{and}
  \bibinfo{author}{\bibfnamefont{W.~J.} \bibnamefont{Marciano}},
  \bibinfo{journal}{Phys. Rev.} \textbf{\bibinfo{volume}{D30}},
  \bibinfo{pages}{248} (\bibinfo{year}{1984}).

\bibitem[{\citenamefont{Djouadi}(2008)}]{Djouadi:2005gi}
\bibinfo{author}{\bibfnamefont{A.}~\bibnamefont{Djouadi}},
  \bibinfo{journal}{Phys. Rept.} \textbf{\bibinfo{volume}{457}},
  \bibinfo{pages}{1} (\bibinfo{year}{2008}), \eprint{hep-ph/0503172}.

\bibitem[{\citenamefont{Cao et~al.}(2015)\citenamefont{Cao, Wang, and
  Zhang}}]{Cao:2015iua}
\bibinfo{author}{\bibfnamefont{Q.-H.} \bibnamefont{Cao}},
  \bibinfo{author}{\bibfnamefont{H.-R.} \bibnamefont{Wang}}, \bibnamefont{and}
  \bibinfo{author}{\bibfnamefont{Y.}~\bibnamefont{Zhang}},
  \bibinfo{journal}{Chin. Phys. C} \textbf{\bibinfo{volume}{39}},
  \bibinfo{pages}{113102} (\bibinfo{year}{2015}), \eprint{1505.00654}.

\bibitem[{\citenamefont{Alwall et~al.}(2014)\citenamefont{Alwall, Frederix,
  Frixione, Hirschi, Maltoni, Mattelaer, Shao, Stelzer, Torrielli, and
  Zaro}}]{Alwall:2014hca}
\bibinfo{author}{\bibfnamefont{J.}~\bibnamefont{Alwall}},
  \bibinfo{author}{\bibfnamefont{R.}~\bibnamefont{Frederix}},
  \bibinfo{author}{\bibfnamefont{S.}~\bibnamefont{Frixione}},
  \bibinfo{author}{\bibfnamefont{V.}~\bibnamefont{Hirschi}},
  \bibinfo{author}{\bibfnamefont{F.}~\bibnamefont{Maltoni}},
  \bibinfo{author}{\bibfnamefont{O.}~\bibnamefont{Mattelaer}},
  \bibinfo{author}{\bibfnamefont{H.~S.} \bibnamefont{Shao}},
  \bibinfo{author}{\bibfnamefont{T.}~\bibnamefont{Stelzer}},
  \bibinfo{author}{\bibfnamefont{P.}~\bibnamefont{Torrielli}},
  \bibnamefont{and} \bibinfo{author}{\bibfnamefont{M.}~\bibnamefont{Zaro}},
  \bibinfo{journal}{JHEP} \textbf{\bibinfo{volume}{07}}, \bibinfo{pages}{079}
  (\bibinfo{year}{2014}), \eprint{1405.0301}.

\bibitem[{\citenamefont{Tanabashi et~al.}(2018)}]{Tanabashi:2018oca}
\bibinfo{author}{\bibfnamefont{M.}~\bibnamefont{Tanabashi}}
  \bibnamefont{et~al.} (\bibinfo{collaboration}{Particle Data Group}),
  \bibinfo{journal}{Phys. Rev.} \textbf{\bibinfo{volume}{D98}},
  \bibinfo{pages}{030001} (\bibinfo{year}{2018}).

\bibitem[{\citenamefont{Brivio et~al.}(2019)\citenamefont{Brivio, Corbett, and
  Trott}}]{Brivio:2019myy}
\bibinfo{author}{\bibfnamefont{I.}~\bibnamefont{Brivio}},
  \bibinfo{author}{\bibfnamefont{T.}~\bibnamefont{Corbett}}, \bibnamefont{and}
  \bibinfo{author}{\bibfnamefont{M.}~\bibnamefont{Trott}},
  \bibinfo{journal}{JHEP} \textbf{\bibinfo{volume}{10}}, \bibinfo{pages}{056}
  (\bibinfo{year}{2019}), \eprint{1906.06949}.

\end{thebibliography}

\end{document}